# Vesper: Using Echo-Analysis to Detect Man-in-the-Middle Attacks in LANs

Yisroel Mirsky, Naor Kalbo, Yuval Elovici, and Asaf Shabtai
Department of Software and Information Systems Engineering,
Ben-Gurion University, Beer-Sheva, Israel.
yisroel@post.bgu.ac.il, kalbo@post.bgu.ac.il, elovici@bgu.ac.il, and shabtaia@bgu.ac.il

***Abstract*—The Man-in-the-Middle (MitM) attack is a cyber-attack in which an attacker intercepts traffic, thus harming the confidentiality, integrity, and availability of the network. It remains a popular attack vector due to its simplicity. However, existing solutions are either not portable, suffer from a high false positive rate, or are simply not generic.**

**In this paper, we propose *Vesper*: a novel plug-and-play MitM detector for local area networks. *Vesper* uses a technique inspired from impulse response analysis used in the domain of acoustic signal processing. Analogous to how echoes in a cave capture the shape and construction of the environment, so to can a short and intense pulse of ICMP echo requests model the link between two network hosts. *Vesper* uses neural networks called autoencoders to model the normal patterns of the echoed pulses, and detect when the environment changes. Using this technique, *Vesper* is able to detect MitM attacks with high accuracy while incurring minimal network overhead.**

**We evaluate *Vesper* on LANs consisting of video surveillance cameras, servers, and PC workstations. We also investigate several possible adversarial attacks against *Vesper*, and demonstrate how *Vesper* mitigates these attacks.**

## I. Introduction

A Man-in-the-Middle attack (MitM) is where a malicious third party takes control of a communication channel between two or more endpoints by intercepting and forwarding the traffic in transit. An attacker in the middle has the capability of harming the confidentiality, integrity, and availability of the user's content, by eavesdropping, manipulating, crafting, and dropping traffic on the network. In general, the MitM attack model on a local area network (LAN) has three steps: (1) gain access to the network, (2) intercept traffic in transit, and (3) manipulate, craft, or drop traffic.

Depending on the scenario, access can be achieved by connecting to a public Wi-Fi access point (e.g. at a café, airport...) or by connecting physically to an exposed network cable or network switch. The attacker can also conduct this attack remotely via a malware which has infected a trusted computer within the existing network [1]. After gaining access, interception can be achieved by exploiting known vulnerabilities in network protocols. For example, the attacker can poison a host's address resolution protocol (ARP) table to capture local traffic [2]–[4], or spoofing a domain name server (DNS) to intercept all web traffic [5]–[7]. The attacker can easily exploit these vulnerabilities with free tools which work out-of-the-box such as Ettercap, Cain and Abel, Evilgrade, arpspoof, dsniff, and many others.

Although MitM attacks on LANs have been known for some time, they are still considered a significant threat [8], [9], and have gained academic attention over the years. This is likely because the attack is relatively easy to achieve, yet challenging to detect [10]. Encryption can protect the integrity and confidentiality of the traffic in transit. However, according to [11], 30% of the world's web traffic is not encrypted. Furthermore, in many cases networked systems do not encrypt their traffic by default (e.g., SCADA control system [12]). Moreover, even if the traffic is encrypted, encryption protocols may have flaws [13], [14], be misconfigured, or simply left out by a manufacturer (e.g. CVE-2017-15643). We also note that LAN-based MitM attacks are used in APTs to achieve lateral movement [15]. Therefore, there is a need for detecting the presence of a MitM, even when encryption is employed.

### A. The Proposed Solution

Our proposed solution is inspired by signal processing domain. In a dynamic system, the output (reaction) of the system to a short input signal is called impulse response. A common use of impulse responses is the modeling and recreation of acoustic environments, such as small rooms or concert halls. As an intuitive example, one can hear the IR of a room by clapping their hands. The sound of the clap changes based on the size, shape, and materials of the room.

Using this concept, we propose a MitM detector called *Vesper*. Vesper bats are the largest and best-known family of the bat species. Akin to it name, our detector captures the impulse response of a LAN by measuring the round-trip-times (RTT) resulting from a short intense burst of ICMP echo requests. This impulse response is used to model the normal behavior of the network in the perspective of two communicating hosts. When a third party intercepts traffic, the harmonic composition of the impulse response between the hosts changes significantly. *Vesper* detects this change using an autoencoder neural network as an anomaly detector. In this paper, we show how *Vesper* works with various devices, in the presence of diverse traffic, and across multiple switches. We also show how *Vesper* is robust against adversarial attacks.

### B. Contributions

To summarize, the contributions of this paper are as follows.

- A novel method for detecting the presence of a MitM attack on a LAN via echo-analysis. The method is non-intrusive (no packet inspection), incurs a minimal overhead on the network, and is not dependent of the hardware and software of the LAN or the attacker's device.
- A framework for deploying the technique on a LAN (*Vesper*). The framework is plug-and-play, which makes the detection method practical. The framework has been designed to resist an evasive adversary.

The rest of this paper is organized as follows. In sections II and III, we present the MitM attack model and related works respectively. In section IV, we provide a background on echo analysis, and introduce our technique (ping echo analysis). In section V, we present the framework for MitM detection in LANs via ping echo analysis (*Vesper*). In section VI, we present evaluations of *Vesper* on several different networks. In section VII, we present possible adversarial attacks against *Vesper* and the respective countermeasures. Finally, in section VIII, we conclude our paper.

## II. The Attack Model

In this section we describe the MitM attack model used throughout the paper. We also enumerate the attacker's requirements, attack vectors, and capabilities.

### A. Attack Scenario

Let Alice and Bob be victims located on the same LAN segment, where the LAN segment may contain one or more network switches. Let Eve be the attacker whose objective is to perform a MitM attack between Alice and Bob. In other words, Eve wants to manipulate the traffic sent between Alice and Bob, while being able to craft new traffic as well (e.g., sending ARP packets). Eve has physical access to the LAN's infrastructure, and can install malware on a network host other than Alice and Bob.

### B. Attack Topologies

Eve can accomplish her objective by establishing one of the following MitM topologies (illustrated in Fig. 1):

**End-Point (EP) MitM.** Eve either adds a new host, or compromises an existing host on the LAN. Eve then causes the traffic in transit between the Alice and Bob to flow through her device (e.g., via ARP poisoning or some other protocol-based MitM attack).

**In-Line (IL) MitM.** Eve locates an exposed network cable which Alice and Bob use to communicate. Eve then covertly installs a device which passes all traffic from one side of the wire to the other, while being able to manipulate/inject traffic.

**In-Point (IP) MitM.** Eve locates an exposed network switch which Alice and Bob use to communicate. Eve then swaps the switch with a new switch that has additional logic enabling her to manipulate/inject traffic.

Unlike the EP MitM, IL and IP MitM attacks can only be accomplished by introducing additional hardware. Therefore, these attacks require physical access to the LAN.

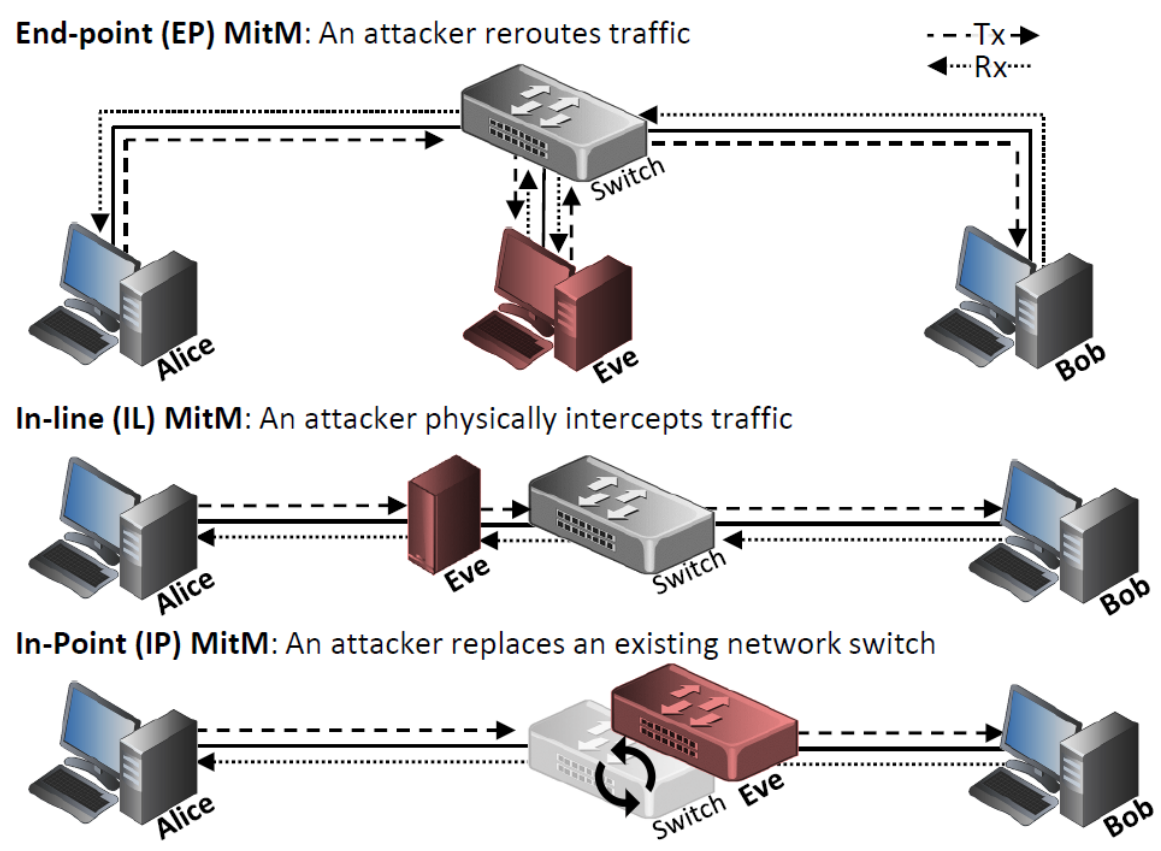

Fig. 1: The three LAN MitM attack topologies.

### C. Classes of Attacks

MitM attacks in a LAN vary based on their stealth and complexity. For example, a more covert attack is typically more difficult for the attacker to accomplish. We categorize the class of a MitM attack based on the MitM topology, and implementation used. Table I summarizes these classes, and their notations which we use throughout the paper.

We note that although an IP-DH MitM is very hard to detect, it is also very hard for the attacker to accomplish. This is because (1) network switches are typically stored under lock and key, and (2) modern switches provide a password protected administrator console (the attacker must copy the configurations prior to the swap).

There are two reasons why such a MitM will buffer each and every inbound packet: (1) to avoid signal collisions on the media when transmitting crafted/altered packets, and (2) to capture and alter relevant packets before they reach their intended destination. In the latter case, the MitM must parse every frame in order to determine the frame's relevancy to the attack, and cannot retroactively stop a transmitted frame. Therefore, the interception process (hardware and/or software) will affect the timing of network traffic. We note that since passive wiretaps only observe traffic, they are not MitM attacks and therefore not in the scope of this paper. However, *Vesper* can detect a MitM which is presently eavesdropping (not currently altering traffic) because a MitM always buffers each packet upon reception. Fig. 2 illustrates the basic packet interception process for all MitM implementations.

## III. Related Works

In our review of related works, we will focus on MitM attacks which the target communication channels [9], and we will also only focus on solutions which are concerned with the data link layer of the OSI model.

In the past decade, many different detection schemes have been proposed in order to address MitM attacks. In general, the solutions to MitM attacks on LANs address a specific flaw in a protocol [3], [5], [6], [16], [17]. As an example, consider the infamous vulnerability in the Address Resolution Protocol (ARP). The vulnerability gives untrusted hosts the ability to spread spoofed ARP messages, causing network traffic to be

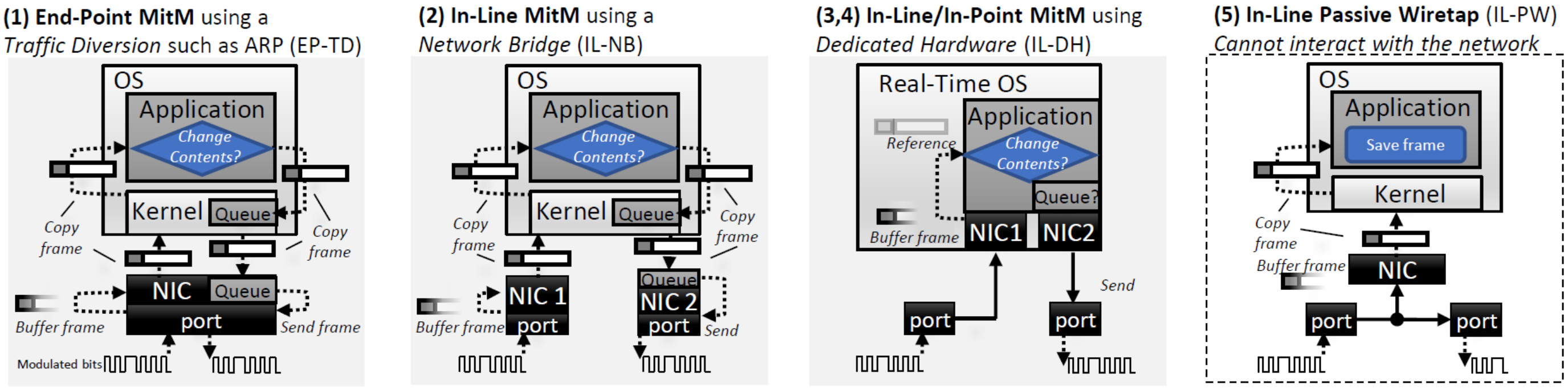

Fig. 2: The packet interception process for (1) an EP MitM, (2-3) an IL MitM, (4) an IP MitM, and (4) a passive wiretap.

TABLE I: Summary of MitM Attacks on a LAN

**TD**: Traffic Diversion **EP**: End-Point
**NB**: Network Bridge **IL**: In-Line
**DH**: Dedicated Hardware **IP**: In-Point
- low, ○ medium, ● high, ★ very-high

| | | MitM 1 | MitM 2 | MitM 3 | MitM 4 | Wiretap 5 |
|---|---|---|---|---|---|---|
| **Type** | *Topology* | EP | IL | IL | IP | IL |
| | *Implementation* | TD | NB | DH | DH | DH |
| **Attack Vectors** | *Gain physical access to switch* | X | X | X | X | X |
| | *Connect to Ethernet wall socket* | X | | | | |
| | *Install malware on existing host* | X | | | | |
| | *Install device on strategic wire* | | X | X | | X |
| **Traffic Intercepted** | *All traffic in LAN* | X | | | | |
| | *Local traffic only* | | X | X | X | X |
| Can the MitM alter and inject traffic? | | X | X | X | X | |
| Complexity of the MitM attack | | - | ○ | ● | ★ | ○ |
| Vesper's detection of the MitM attack | | ★ | ★ | ● | ○ | NA |

routed to the attacker's device. Solutions to this flaw include improved protocols [2], [18], [19] and the integration of new security features [20].

Intrusion Detection Systems (IDS) have been proposed as a more generic way for dealing with MitM attacks. These IDSs include software-based IDSs [21] and hybrid hardware/software IDSs (e.g., an add-on component plugged into the mirror port of a switch) [10].

However, these solutions have limitations:

**Generalization.** Many of these solutions address a flaw in a specific protocol, and therefore cannot be generalized to other or unknown MitM attacks occurring in the LAN. For example, detecting an exploitation of the ARP protocol does not solve the issue of an IL MitM.

**Portability.** Some of these solutions require additional hardware or other costly resources. For example, a separate network host which acts as an IDS.

**False Positives.** Network traffic tends to be noisy, making it difficult to detect the presence of a MitM based on traffic patterns and packet contents. Therefore, searching network packets for evidence of a MitM may lead to a large number of false positives [10].

Regardless, all of the related solutions are weak to IL MitM attacks, since they leave no forensic evidence in the packets. On the other hand, *Vesper* is portable and can detect both EP and IL MitM attacks. Furthermore, *Vesper* is robust since it analyzes its own probes and not the traffic of others.

In [22], the authors propose a method for detecting when an end-user's Network Interface Card (NIC) has been changed to promiscuous mode (e.g., sniffing via Wireshark). They accomplish this by creating an RTT dataset for each host in the network, categorized by the host's operating system. Each dataset contains the average, standard deviation, and ratio between the statstics with and without the sniffer active on the end-point. The authors then use z-statistics over these datasets to determine if a host has activated a sniffer.

Our work differs from [22] in the following ways:

1) The technique in [22] detects hosts operating in promiscuous mode, but not necessarily acting as a MitM.
2) The solution in [22] requires the network administrator to *manually* collect a data set from each device in his/her network; with *and* without promiscuous mode on. Our technique is transparent, and provides a fully unsupervised plug-and-play solution.
3) The solution in [22] is weak against replay attacks. Our solution is robust against these attacks because we utilize features extracted from randomized signals.

## IV. Echo Analysis

In this section, we present the probing technique used by *Vesper* to capture the presence of a MitM. Later in section V, we show how *Vesper* uses this technique to actively detect MitM attacks.

### A. Notations and Definitions

In order describe our technique, we now briefly present some common notations and definitions from the domain of signal processing [23].

**Signal.** A signal, in this paper, is a discreet sequence of values sampled at the rate $f_s$ (measured in units of Hz). Let $x$ be a signal where $x[t]$ denotes the value of that signal at time index $t$.

**System.** In signal processing, a system $S$ can be represented as a 'black box' which receives an input signal $x$ and produces an output signal $y$. This process is denoted as $S(x) = y$.

**LTI System.** A special class of systems that are both **l**inear and **t**ime–**i**nvariant. A system is linear if (1) it obeys the additivity principal in that the input $x[t] = x_1[t] + x_2[t]$ produces the output $y[t] = y_1[t] + y_2[t]$, and (2) it obeys the homogeneity principle in that for an input $x[t]$ and its output $y[t]$, it holds that the response for $ax[t]$ is $ay[t]$. A system is time-invariant if the output does not depend on *when* the input was applied.

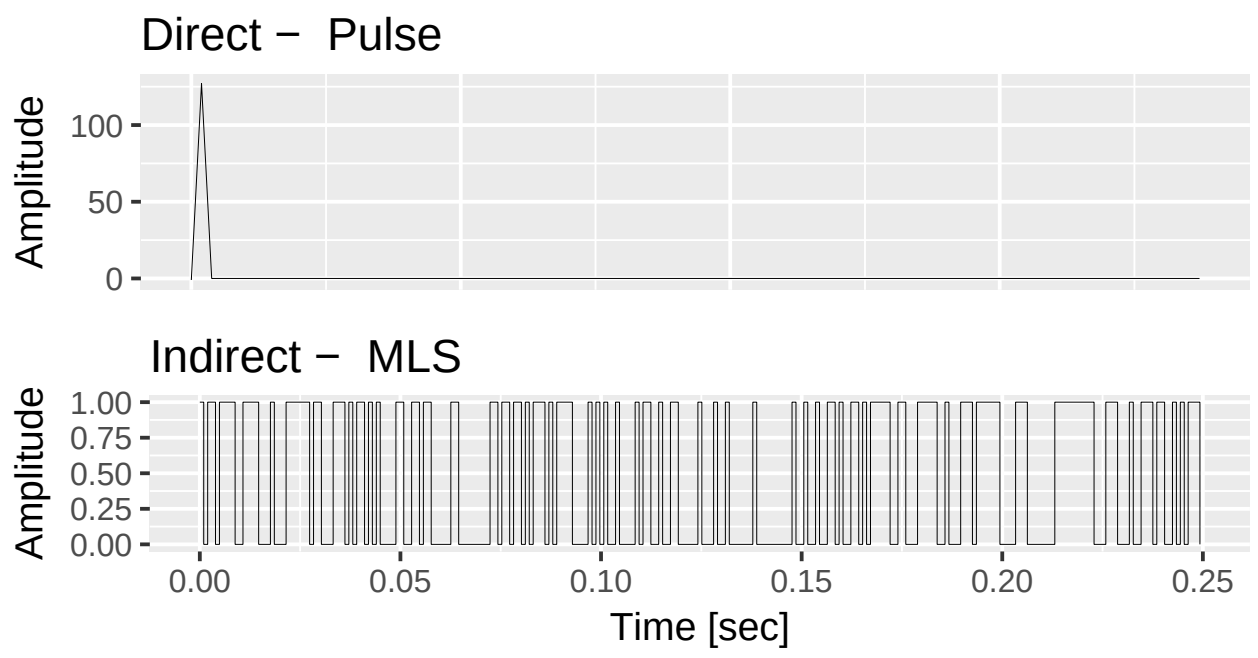

Fig. 3: Two common excitation signals used to extract impulse responses from acoustic environments.

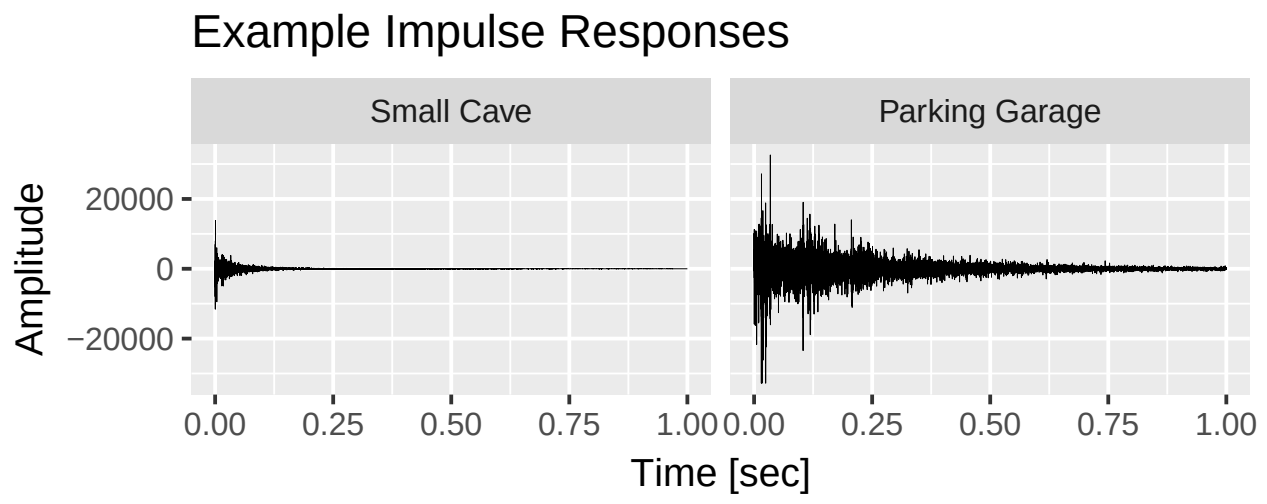

Fig. 4: Example impulse responses extracted from various acoustic environments.

**Excitation Signal.** A brief input signal $x$ which has been crafted to capture the dynamic reaction of $S$ within the output $y$.

**Impulse Response.** An impulse response signal $h$ of the LTI system $S$ is the output of $S$ when presented with a brief excitation signal called an impulse (the Dirac delta function). Since the impulse signal contains all frequencies, $S$ is completely characterized by $h$. This means that for *any* input $x$, the output of $h$ can be calculated as the convolution $y[t] = h[t] * x[t]$.

### B. Acoustic Signal Processing

Our technique is inspired by the domain of acoustic signal processing. Therefore, we will now briefly cover this topic to give the reader a better understanding of our technique.

In the domain of acoustic signal processing, a sound which reverberates through the air, and the environment (e.g., room) which reflects and affects the vibrations as they propagate, are the signal $x$ and LTI system $S$ respectively. An acoustic engineer can model $S$ by extracting its impulse response $h$. This can be achieved by emitting an excitation signal $x$ at one location while simultaneously recording the resulting signal $y$ at another location. In this case, the input to $S$ is generated by a speaker and the output is captured by a microphone. There are several methods for extracting an acoustic impulse response with an excitation signal. These methods can be categorized as either being direct or indirect (visualized in Fig. 3):

**Direct Methods.** Direct methods involve an excitation signal $x$ which is similar to that of a Dirac function, so that $y = h$. However, since it is impossible to generate a true Dirac signal in an acoustic environment, short loud sounds are used instead. For example, popping a balloon, generating a spark, and firing a gun.

**Indirect Methods.** An approximation of $h$ can be obtained *indirectly* from a non-Dirac excitation signal. The process involves deconvolving the excitation signal $x$ with the resulting output signal $y$ [23]. One well-known excitation signal is the maximal length sequence (MLS) signal. An MLS is a pseudorandom binary sequence generated from maximal linear feedback shift registers. With $m$ registers, the generator produces a random binary sequence of length $N = 2^m - 1$ which is spectrally flat. As a result, an MLS excitation signal produces all frequencies, closely resembles white noise, and is robust in noisy and populated environments [24].

Once the impulse response $h$ has been extracted from $S$, it can be used to perform a convolution reverb (a digital simulation of an audio environment on sound). For example, the response can be used to make a recorded piece of music to sound like it was played in a particular cave or arena. We can see from this that $h$ is dependent on the shape of the room, the materials of the surfaces, and the positioning of the speaker and microphone. Any alteration to these physical parameters will cause a noticeable affect on the impulse response. In other words, an impulse response can be seen as an *acoustical signature* of the environment.

To illustrate this concept, Fig. 4 presents two impulse responses extracted from different environments. The initial Dirac pulse (e.g., balloon pop sound) can be seen at the beginning, followed by dynamic reverberations and echoes (i.e., spikes). The figure shows that each environment has its own unique signature due to their unique constructs.

### C. Ping Signal Processing

Our approach to MitM detection is to (1) model a LAN as an acoustic environment, (2) emit excitation signals, (3) model the echoed response signals, and (4) detect abnormal changes in newly sampled responses.

In networks, there are no reverberations of sound waves. However, switches, network interfaces, and operating systems all affect a packet's travel time across a network. The hardware, buffers, caches, and even the software versions of the devices which interact with the packets, all affect packet timing. This is analogous to how a sound wave is affected as it reverberates off various surfaces.

To capture packet timing between a local host and an end-host, one can use the Internet Control Message Protocol (ICMP) [25]. The ICMP is a popular protocol used to gain feedback about problems in an IP network. One of the features of this protocol is the `Echo_Request` command, commonly used to determine whether a host is operational. When a host sends another host an `Echo_Request`, the target host returns an ICMP `Echo_Reply`. Upon receiving the `Echo_Reply`, the sender can measure the round-trip-time (RTT) to and from the receiver. This process is referred to as 'pinging'. To according to the ICMP standard (RFC 1122: 3.2.2.6), one may include data (a payload) in an `Echo_Request`. In this case, the receiver must include the same data in the `Echo_Reply`.

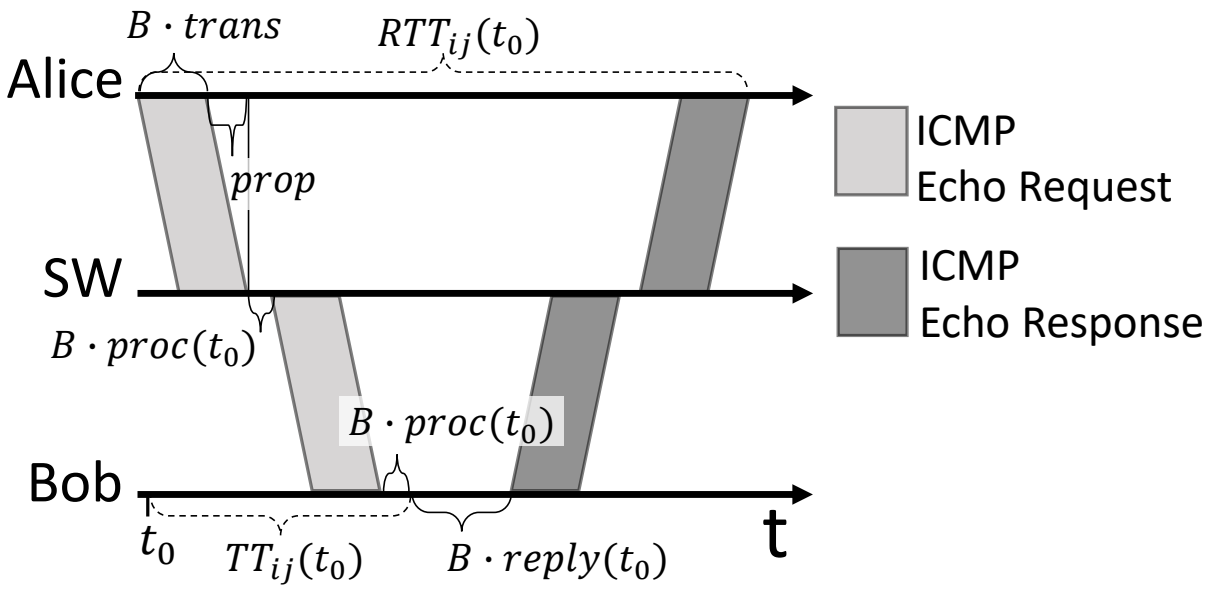

Fig. 5: An illustration of the ICMP RTT model. The propagation time has been exaggerated for visualization purposes.

The RTT of an ICMP packet over a LAN is dependent on the number of switches (hops) traversed, since interactive networking elements (e.g., switches) must fully buffer each received frame before transmitting. The RTT is also dependent on the current load and the hardware/software implementation of each networking element along the path of the ping. With this in mind, we present the following RTT model (illustrated in Fig. 5):

Let $h_i(t)$ denote the time it takes to transfer an Ethernet frame between the two network elements at the $i$-th hop at time $t$. Assuming the frame is $B$ bytes long, we model it as

$$h_i(t) = prop + B\,(trans + proc_i(t)) \tag{1}$$

where $prop$ is the propagation time on the wire (approx. the speed of light), $trans$ is the transmission time for a single byte, and $proc_i(t)$ is the average processing time per byte at the transmitting element at time $t$ (e.g., parsing and buffering).

For simplicity, we will assume that all switches in the LAN are identical. As a result, using (1) we can model the trip time (TT) from host $i$ to host $j$ as

$$TT_{ij}(t) = h_1(t) + h_2(t + h_1) + \ldots + h_k(t + h_1 + \ldots + h_{k-1}) \tag{2}$$

Finally, we can use (2) to model the RTT between host $i$ and host $j$, sent at time $t$, as

$$RTT_{ij}(t) = TT_{ij}(t) + B \cdot reply_j(t) + TT_{ji}(t + TT_{ij}) \tag{3}$$

where $reply_j(t)$ is the average time it takes for host $j$ to process each byte in an ICMP echo reply, at time $t$.

With the assistance of the model in (3), we will now define our system and its signals.

*1) System Definition* ($S$)

Let $S$ be a LAN environment consisting of one or more switches and numerous hosts. Let $S_{ij}$ be the LAN in the perspective of host $i$ communicating with host $j$, where $i$ and $j$ are within the same LAN.

We define the input signal $x$ to $S$ as a sequence of ICMP `Echo_Request` frames, where $x[n] \in \{42, 43, \ldots, 1542\}$ are the number of bytes which are transmitted in the ICMP `Echo_Request`: 42 bytes for the Ethernet, IPv4, and ICMP protocol headers, plus an additional 0-1500 bytes for the ICMP payload). We define the output signal $y$ as a sequence of RTTs, computed from the respective ICMP `Echo_Reply` packets' arrival times. More formally,

$$y[n] = T_{rx}[n] - T_{tx}[n] \tag{4}$$

where $T_{tx}[n]$ is the transmission timestamp of the $n$-th `Echo_Request` in $x$, and $T_{rx}[n]$ is the reception time of the resulting `Echo_Reply`.

When the random sized requests in $x$ are sent back-to-back at a fast rate, the electronics, caching mechanisms, CPU schedulers, and queuing algorithms of each network element dynamically affect the respective $proc(t)$ and $reply(t)$ in response to the varying load. Since the payloads in $x$ reflect an MLS signal, $y$ captures $S_{ij}$'s fingerprint (impulse response).

Empirical evidence can be shown via linear regression. In brief, we found that the $k$-th RTT in $y$ has a dependency on the random sizes of previously transmitted ICMP requests in $x$. More formally, $y[k] \sim x[1], x[2], \ldots, x[k]$. We also found that an 18% reduction in error can be achieved by considering $x[1], \ldots, x[k]$ as the descriptor variables as opposed to just $x[k]$. An example of this dependency can be visualized in Fig. 7. The figure plots the RTT distribution of the $i$-th ping in a back-to-back burst of 50 pings (sent 1500 times). If there were no dependency, then the distribution of the $i$-th and $j$-th ping would be identical.

Fig. 7 also shows that the first pings are noisier than those which follow (e.g., due to caching). This is another reason why we must send $x$ at a fast rate, and not as individual pings. In our system, we set the transmission rate of $x$ to

$$f_s \equiv \frac{2}{\mu_{RTT^*}} \tag{5}$$

where $\mu_{RTT^*}$ denotes the average RTT time of largest ping possible (a 1542 byte Ethernet frame). This rate ensures that $y$ captures the system well, while not overloading the end-host.

*2) Ping Excitation Signal* ($x$)

*In order to capture a characterization of $S_{ij}$, we use an MLS excitation signal as our input $x$.*

To transmit a binary MLS, we modulate the sequence over the minimum and maximum ICMP payload sizes. For example, one possible $N = 7$ length MLS may be $s = \{1, 1, 1, 0, 1, 0, 0\}$. In this case, $s$ would be translated into the transmission signal $x = \{1542, 1542, 1542, 42, 1542, 42, 42\}$. Fig. 6 illustrates an $m = 10$ bit ($N = 1023$ length) sequence modulated as the input signal $x$, and then received as the output signal $y$.

There are several reasons why we use the MLS method over other known excitation methods:

- The MLS method is known to be robust in noisy environments, such as a room populated with conversing individuals [24]. Network traffic can affect $S_{ij}$, thus it is appropriate to assume the system will be noisy.
- An MLS of sufficient length has subsequences of '1's. This results in bursts of pings which have the maximum size of 1500. This burst causes a momentary stress on the network which is reflected in the output $y$, thus better capturing the network's characteristics.

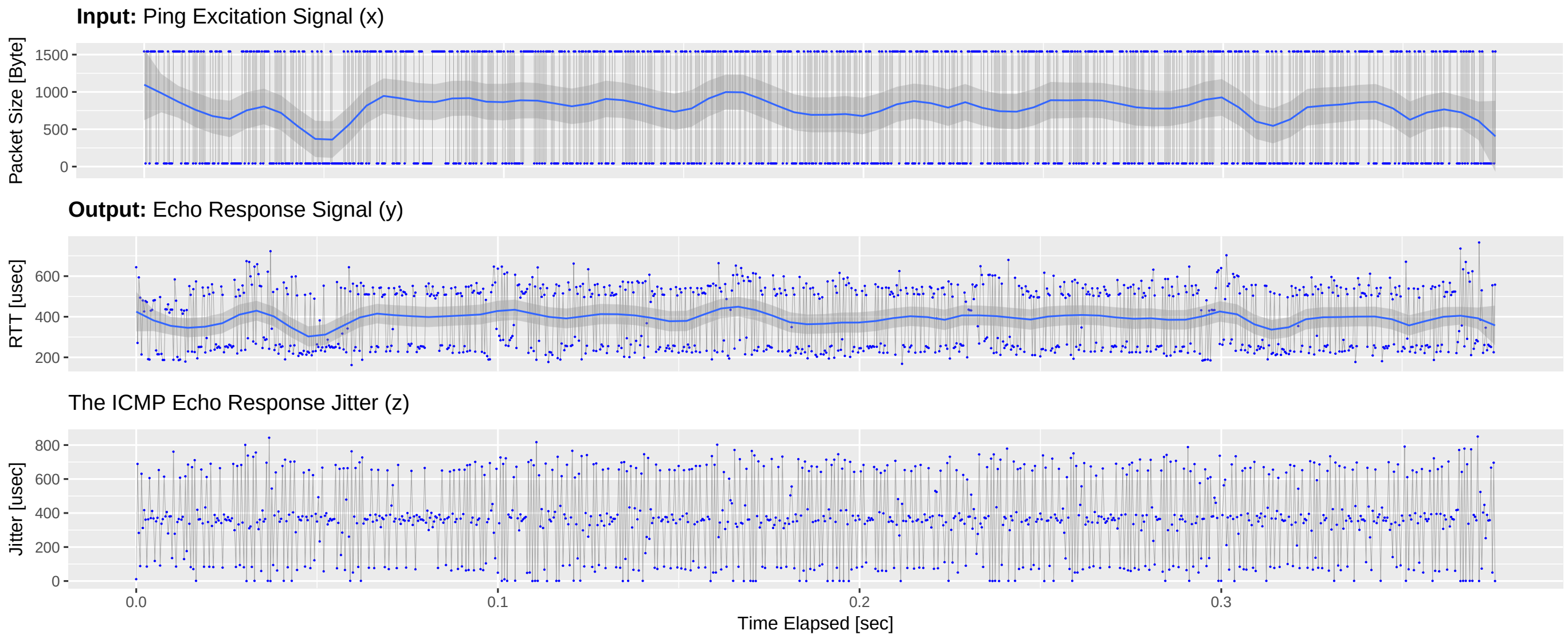

Fig. 6: Sample signals from a Linux PC ($i$) and an IP-based security camera ($j$). Top: the 10 bit MLS ping excitation signal $x$, sent from $i$ to $j$. Middle: the resulting ping echo response signal $y$, timed by $i$. Bottom: the ICMP echo response jitter $z$, computed by $i$. Signals $x$ and $y$ have been fitted to a Loess curve (blue line) for better visualization.

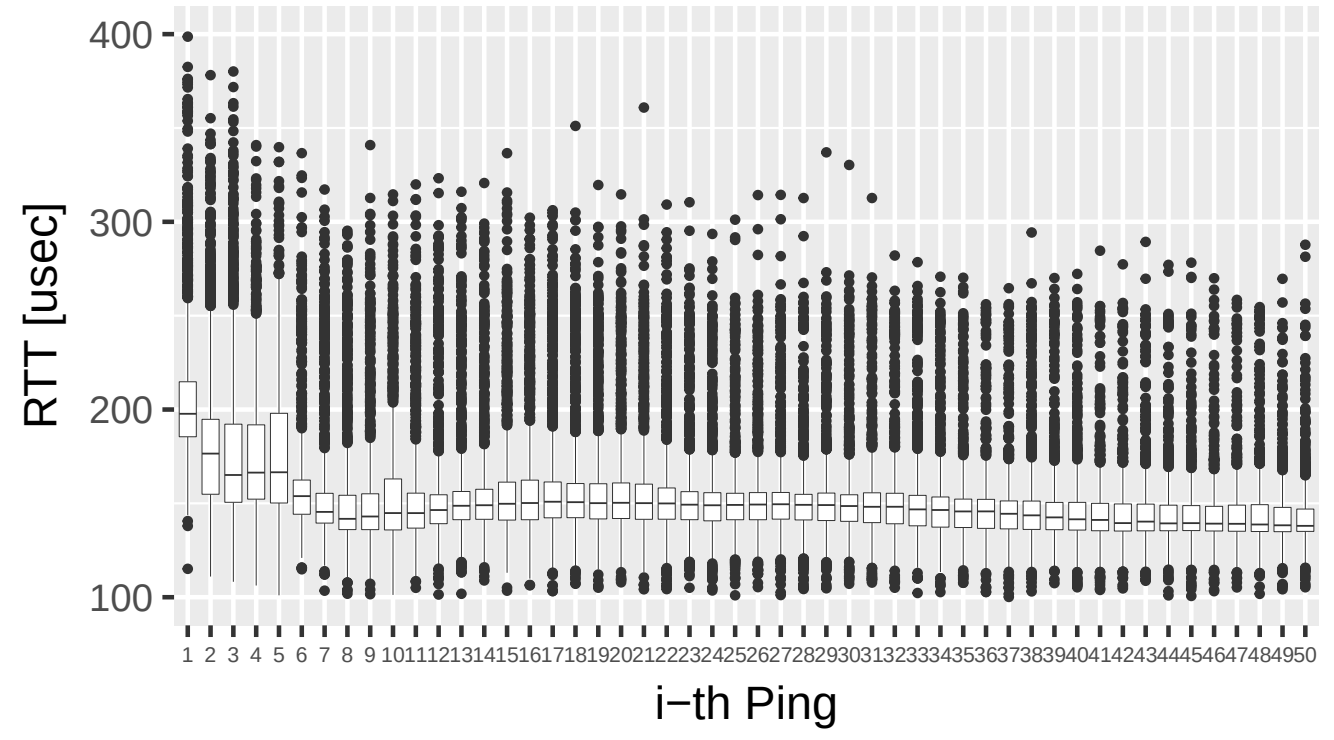

Fig. 7: The distribution of each ping's RTT, when sending a burst of 50 pings back-to-back. The data was collected over 1500 trials with a direct host-to-host Ethernet connection.

- An MLS is randomly generated each time, thus raising the difficulty for an attacker to perform a replay attack (discussed in detail later in section VII).

*3) Echo Response Signal* ($y$)

*The system's output y, from the excitation signal x, contains the impulse response h.*

In order for $h$ to fully characterize $S_{ij}$, the system $S_{ij}$ must be LTI. However, since RTTs are dependent the media's propagation time ($prop$), there is non-linear relationship and $S_{ij}$ does not abide to the homogeneity principal. Therefore, strictly speaking, $S_{ij}$ is not LTI.

However, when modeled over a short interval as a snapshot of the network, $S_{ij}$ can be described as an LTI system, and it's impulse response can significantly capture the system's characteristics at that moment:

$S_{ij}$ is a non-linear system because the propagation time $prop$. However, since $prop$ is close to the speed of light, this term is negligible with respect to the rest of the timing model in (3). Therefore, $prop$ has no significant affect on the system, and $S_{ij}$ can be viewed as a linear system.

$S_{ij}$ is time-variant because $proc(t)$ and $reply(t)$ are dependent on the load of the LAN and the end-host, which changes over time. Let $S_{ij}^{(t)}$ be the state of the system $S_{ij}$ at time $t$. Let $x^{(t)}$ and $y^{(t)}$ be the respective input and output of the system at time $t$. If $x^{(t)}$ is significantly short, then processing times of $S_{ij}^{(t)}$ can be approximated as constants. Therefore, snapshots of $S_{ij}$ are independently time-invariant.

In summary, $S_{ij}$ cannot be characterized with a single response to an MLS signal. However, by sampling the distribution of $S_{ij}$'s impulse responses, we can implicitly capture $S_{ij}$'s characterization over time. Therefore, a collection of echo responses can be used to model $S_{ij}$'s normal behavior (see section V-E).

*4) ICMP Echo Response Jitter* ($z$)

Jitter is the time lapse between two consecutive packet arrivals. We denote the jitter values of the ICMP echo responses as $z$, defined as

$$z[t] = T_{rx}[t] - T_{rx}[t-1] \tag{6}$$

The bottom of Fig. 6 plots an example of the jitter resulting from the MLS signal $x$. In this example, we can see that the jitter is distributed at distinct high (650 usec), medium (390 usec), and low (50 usec) levels. The three distinct levels are the result of the transitions between adjacent bits in the MLS binary sequence. For instance, whenever a '10' appears in the sequence, the jitter is small. This is because the RTT of a 42 byte packet ('0') is shorter than that of a 1542 byte packet ('1'). Since the pings are sent at a rate of $f_s$, the response for the '0' arrives shortly after the response for the '1'.

Although $z$ is not part of our system model $S_{ij}$, it captures additional characteristics of the channel between $i$ and $j$. For example, additional processing delays and moments of stress on the participating network elements.

## V. Vesper

In this section, we present the MitM detector *Vesper*: the framework, machine learning process, and deployment.

### A. Overview

*Vesper* is a *plug-and-play* man-in-the-middle detector based on ping echo-analysis. The detector is installed on a local host within a LAN, and protects the local host from MitM attacks originating from within the same LAN. In this section, we use $\Gamma$ to denote the set of known remote hosts in the LAN, excluding the local host.

*Vesper*'s framework has the following main components:

- **Orchestrator (OR):** The component responsible for adding new local IPs (hosts) automatically, and deciding which link in the LAN should be probed when.
- **Link Prober (LP):** The component responsible for probing the hosts in $\Gamma$. Each probe produces an MLS excitation signal ($x$), which results in the echo response signal ($y$) and the echo response jitter ($z$).
- **Feature Extractor (FE):** The component responsible for summarizing the result of a probe. The summary forms a feature vector $\vec{v} \in \mathbb{R}^3$.
- **Host Profiler (HP):** The component responsible for detecting the presence of a MitM using $\vec{v}$. It accomplishes this by profiling the link to each host $j \in \Gamma$ with an autoencoder. The autoencoder is trained to recognize the link's normal behavior. An autoencoder is a neural network which can be used as an anomaly detector (discussed later in section V-E).

*Vesper* operates by performing the following steps (illustrated in Fig. 8):

*Vesper's MitM Detection Procedure*

**I. Orchestrator**

1) At a random time, a random network host $j \in \Gamma$ is selected and the detection process is initiated.

**II. Link Prober**

2) The *MLS Generator* produces the random binary sequence $s$.
3) The *MLS Modulator* creates the ping excitation signal $x$ based on the binary sequence $s$.
4) The *Excitation Emitter* sends host $j$ a total of $N$ `Echo_Request` packets, according to $x$, and at a rate of $f_s$. In parallel, the *Echo Receiver* captures $j$'s `Echo_Reply` packets.
5) Once all $N$ `Echo_Reply` packets have been received, the *MLS Demodulator* extracts the echo response signal $y$ and the echo response jitter $z$.

**III. Feature Extractor**

6) The *Impulse Extractor*, *DC Extractor*, and *KS-Tester* use $x$, $y$, and $z$ to produce the feature vector $\vec{v}$.

**IV. Host Profiler**

7) The IP address of $j$ is used to retrieve $j$'s autoencoder via a hashmap.
8) Using $\vec{v}$, the autoencoder determines whether or not the link with host $j$ has been significantly altered. If $\vec{v}$ is determined to be normal (with high confidence), then the autoencoder learns from the instance $\vec{v}$. Otherwise, an alert is raised.

We will now discuss the each of *Vesper*'s main components in greater detail.

### B. Orchestrator (OR)

Whenever a new IP address from the same subnet as *Vesper* is observed in the network traffic, or added by the user, the OR pings that address. If none of the pings traverse a router (indicated by `TTL` field of the IPv4 header) then the address is added to $\Gamma$. After sending each probe, at a random time within the next second, the OR selects a random host $i \in \Gamma$ and initiates a probe via the LP.

### C. Link Prober (LP)

After generating $s$ and $x$, the LP transmits `Echo_Request` packets to the target host, according to $x$. The `Echo_Request` packets are transmitted every $\frac{1}{f_s} = \frac{1}{2}\mu_{RTT^*}$ seconds. This means that the transmission and reception of ICMP packets must be performed concurrently on two separate threads: the *Excitation Emitter* and *Echo Receiver*.

In order to measure the RTT of each ping correctly, each `Echo_Request` must be associated with its respective `Echo_Reply`. To accomplish this, the *Excitation Emitter* places the current index of $x$ into the `Sequence_Number` field of the `Echo_Request` header. When a host replies, it copies the same value from the `Echo_Request` into the header of its `Echo_Reply`. The `Identifier` field is used to differentiate between separate excitation signals.

In order to obtain the necessary accuracy, the LP records all transmission and reception timestamps with nanosecond resolution. In C++, and with a Linux kernel, this can be accomplished using the `<time.h>` library's `clock_gettime()` with the `CLOCK_MONOTONIC` option enabled.[1]

When the last `Echo_Reply` is received, $y$ and $z$ are computed, and the raw probe data $(x, y, z)$ is passed to the FE for feature extraction.

### D. Feature Extractor (FE)

After each probe, the FE is tasked with extracting the following three features from $(x, y, z)$:

| | |
|---|---|
| $v_{E_h}$: | The impulse response energy using $x$ and $y$ |
| $v_{rtt*}$: | The mean RTT from the largest packets sent in $x$ |
| $v_{jit}$: | The log-likelihood of the jitter's distribution ($z$) |

The feature vector $\vec{v} = \{v_{E_h}, v_{rtt*}, v_{jit}\}$ summarizes the state of the probed channel (system $S_{ij}$). After the FE computes $\vec{v}$, the vector is passed to the HP for inspection. Fig. 10 plots each of the features before and after a MitM attack.

[1]The API instructs the OS to collect the time from a CPU register. On a Dell PC, we found the error to be at worst 0.015% w.r.t. the smallest possible RTT (150 usec with a host-to-host direct Ethernet link).

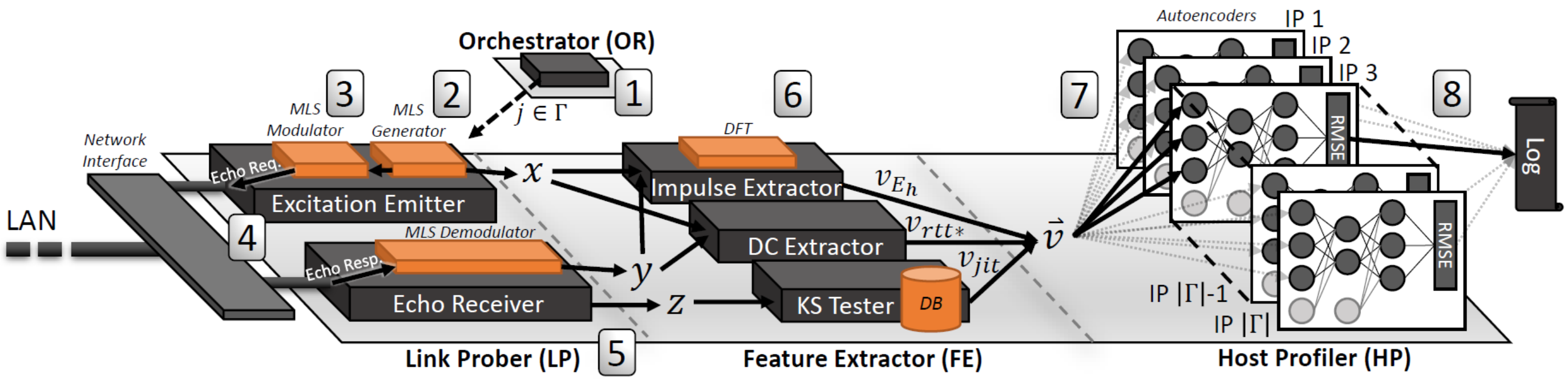

Fig. 8: The framework of *Vesper*, deployed on a network host.

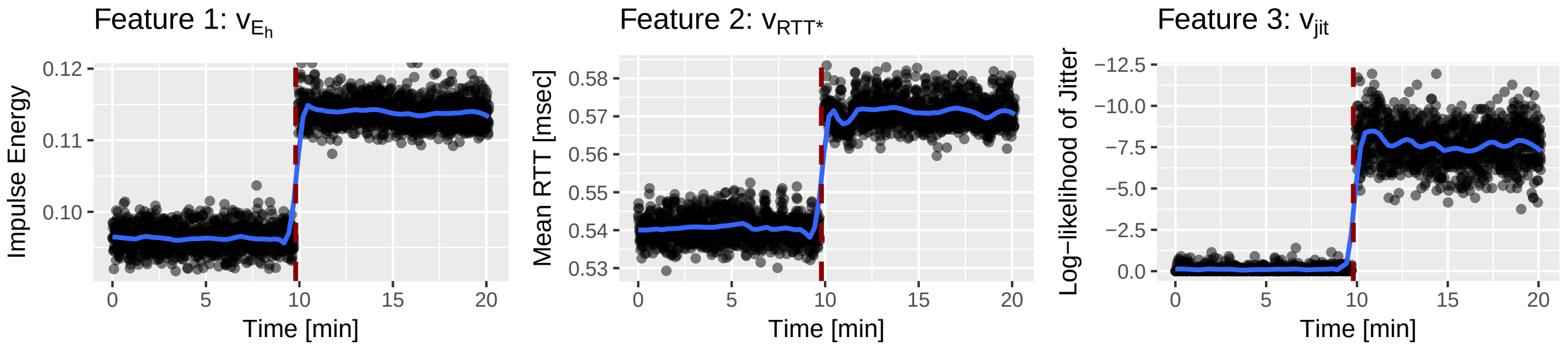

Fig. 9: The three features extracted from probes over time where the dashed line indicates the start of an IL-DH MitM attack.

*1) Impulse Response Energy ($v_{E_h}$)*

When Eve intercepts Alice's traffic, Eve affects the dynamics of the channel between Alice and Bob. Even when Eve responds to Alice's ICMP traffic on behalf of Bob, the impulse response $h$, captured by $y$, changes. This is because Alice has different hardware and software than Bob.

The *Impulse Extractor* summarizes the state of the system $S_{ij}$ with $h$, where $i$ is the local host. It accomplishes this by measuring the energy of $h$, denoted $E_h$. As mentioned in section IV-B, through a process called deconvolution, the response $h$ can be obtained from the output $y$ by knowing the excitation signal $x$.

In an LTI system, the output can be expressed as

$$y[t] = h[t] * x[t] \tag{7}$$

where $*$ is the convolution operator, and $h$ is the system's impulse response. To extract impulse response in the absence of noise, we can perform the deconvolution

$$h[t] = \mathcal{F}^{-1}\{Y/X\} \tag{8}$$

where $\mathcal{F}^{-1}$ is inverse of the Discreet Fourier Transform (DFT), and where $Y$ and $X$ are the Fourier Transforms of the signals $y$ and $x$ respectively.

Using Parseval's theorem [23], we can obtain $E_h$ without the need for computing the inverse DFT in (8) by computing

$$E_h = \frac{1}{N}\sum_{k=1}^{N}\left|\frac{Y[k]}{X[k]}\right|^2 \tag{9}$$

The resulting value used as the feature $v_{E_h}$ in $\vec{v}$.

*2) Mean RTT of the Largest Packets ($v_{rtt*}$)*

In Fig. 1, it can be seen that a both MitM attack scenarios add an additional $2{\cdot}T_{hop}$ delay to Alice's traffic. We can detect this additional delay by averaging the values (RTTs) in $y$.

We note that the duration of $T_{hop}$ increases with the length of the Ethernet frame. Approximately 50% of the packets in $x$ have the maximum length of 1542 bytes. By averaging the RTTs of those frames only, we obtain a better separation between benign the malicious scenarios. Fig. 10 illustrates the benefit of averaging the 1542 byte frames in each response $y$.

This average is extracted from each $y$ by the *DC Extractor*, and used as the feature $v_{RTT*}$ in $\vec{v}$.

*3) Log-likelihood of the Jitter's Distribution ($v_{jit}$)*

As mentioned in section IV-C4, the jitter of the `Echo_Reply` packets ($z$) captures the behavior of the networking elements between the sender and receiver. Concretely, since $x$ is transmitted at a rate of $f_s$, it can be expected that some packets may be being queued, and then transmitted back-to-back. This dynamic behavior characterizes the network's elements, and can be used to help fingerprint the connection with host $j$.

Fig. 11 plots $z$'s distribution, with and without the presence of a MitM attack. Fig. 11 shows that the three levels of jitter (refer to section IV-C4) are affected by the attack. To detect abnormalities in this distribution, the FE performs a two-sample Kolmogorov-Smirnov (KS) test. The KS test is a nonparametric statistical test which results in a probability value (p-value) that indicates how likely two sample distributions come from the same distribution. We denote this value as $p_{X,Y}$, where $X$ and $Y$ are tested distributions.

The *KS Tester* stores $m$ samples of host $j$'s jitter distributions. These samples are used as references for the channel's expected behavior. We denote host $j$'s references as the set $Z_j = z_1, z_2, \ldots, z_m$. Although $m$ is a parameter of *Vesper*, in practice $m = 5$ works well.

Let $z_0$ denote the jitter distribution given to the FE for

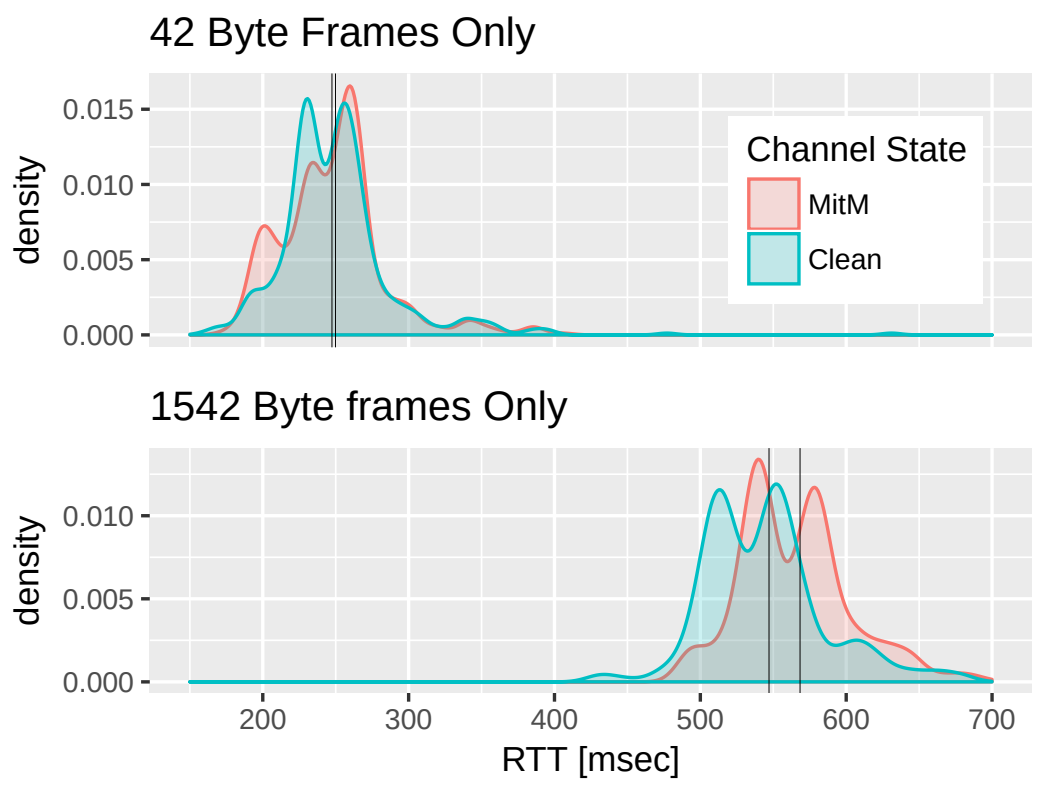

Fig. 10: The distribution of $y$'s RTTs, when exclusively considering the 42 byte or 1542 byte frames, with and without an IL-DH MitM. The bars mark each of the means.

feature extraction. With $z_0$, the *KS Tester* computes the value

$$p_{jit} = \log[max\{p_{z_0,z_1}, p_{z_0,z_2}, \dots p_{z_0,z_m}\}] \tag{10}$$

The last $k$ computations of $p_{jit}$ (from previous probes) are averaged to form the feature $v_{jit}$. In practice, we found that $k = 15$ produces good results. If $v_{jit} \approx 0$, then the *KS Tester* randomly decides whether or not to update $Z_j$ with $z_0$. In (10), we take the maximum p-value, since this makes the feature more robust against false positives.

### E. Host Profiler (HP)

The HP component uses autoencoders to perform the basic anomaly detection. First, we will explain in detail how an autoencoder works, and then we will explain how the HP uses them to detect anomalies in the link with host $j$.

#### 1) Autoencoders

An autoencoder is an artificial neural network (ANN) which is trained to reconstruct it's inputs [26]. During training, an autoencoder tries to learn the function

$$h_\theta(\vec{x}) \approx \vec{x} \tag{11}$$

where $\theta$ is the learned parameters of the ANN, and $\vec{x} \in \mathbb{R}^n$ is an instance (observation). It can be seen that an autoencoder is essentially trying to learn the identity function of the original data distribution. Therefore, constraints are placed on the network, forcing it to learn more meaningful concepts and relationships between the features in $\vec{x}$. The most common constraint is to limit the number of neurons in the inner layers of the network. The narrow passage causes the network to learn compact encodings and decodings of the input instances.

If an instance does not belong to the learned concepts, then we expect the reconstruction to have a high error. The reconstruction error can be computed by taking the root mean squared error (RMSE) between the input $\vec{x}$ and the reconstructed output $\vec{y}$. The RMSE between two vectors is defined as

$$r_{\vec{x},\vec{y}} = \sqrt{\frac{\sum_{i=1}^{n}(x_i - y_i)^2}{n}} \tag{12}$$

where $n$ is the dimensionality of the input vectors.

In order to determine whether or not the observation $\vec{x}$ is an anomaly, we set a cut-off probability $p_{thr}$, and test if $p(X > r_{\vec{x},\vec{y}'}) < p_{thr}$ assuming that $X \sim N(\mu_r, \sigma_r)$, where $\mu_r$ and $\sigma_r$ are sample statistics taken from the benign instances' RMSEs.

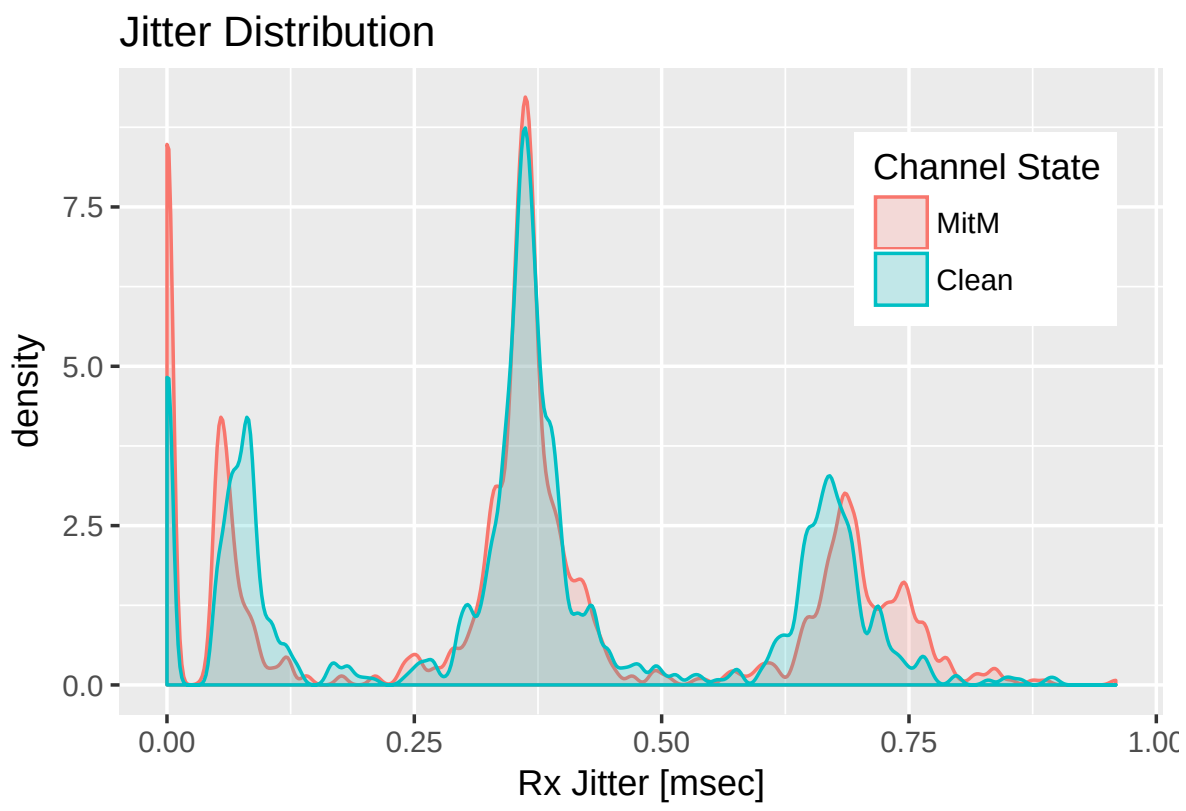

Fig. 11: The distribution of a probe's response jitter ($z$), with and without the presence of an IL-DH MitM.

#### 2) The Anomaly Detection Procedure

To detect anomalies, the HP component maintains an autoencoder for each $j \in \Gamma$ (denoted $A_j$). The task of $A_j$ is to (1) learn the normal behavior of the system $S_{ij}$ via each probe $\vec{v}$ taken from host $j$, and (2) raise an alert if a sample $\vec{v}$ is abnormal. The HP accomplishes this in a plug-and-play fashion by continuously training $A_j$ on non-anomalous data, and by giving $A_j$ a grace period to converge before execution (e.g., only after training on 100 observations). In summary, the HP performs the following steps when instance $\vec{v}$ arrives:

*The Procedure of the HP Component*

1) The model $A_j$ is retrieved via a hashmap, with host $j$ as the key.
2) **`Execute`**$(\vec{v})$: $\vec{v}$ is propagated through $A_j$ to produce the reconstruction $\vec{v}'$.
3) The reconstruction error is computed as $r_{\vec{v},\vec{v}'}$, using $\mu_r$ and $\sigma_r$.
4) **`if`** $p(X > r_{\vec{v},\vec{v}'}) < p_{thr}$, **`and`** the grace period is over, an *alert* is raised.
5) **`else if`** an alert was not detected within the last several probes:[a]
   a) **`Train`**$(\vec{v},\vec{v})$: Using $\vec{v}$, the weights in $A_j$ are updated once using SGD and the back-propagation algorithm [27].
   b) $\mu_r$ and $\sigma_r$ are updated with $r_{\vec{v},\vec{v}'}$.
6) The anomaly score $r_{\vec{v},\vec{v}'}$ is logged.

[a] Reduces the chance of learning from a false negative.

## VI. Evaluation

In this section, we refer to the device on which *Vesper* is installed as *Alice*, and *Bob* as the device whose channel with *Alice* is under attack.

We evaluated *Vesper* in detecting EP-TD, IL-NB, IL-DH and IP-DH MitM attacks (see Table I). For the EP-TD MitM attack, we used a Kali Linux Desktop PC which performed an ARP poising attack. For the IL-NB and IL-DH MitM attacks, we used a Raspberry Pi 3B and a 1Gbps Ethernet switch respectively. The Pi was provided with an extra Ethernet adapter, and configured to operate as a network bridge. For the IP-DH MitM attacks, we used three 1Gbps switches: an advanced feature (SW1), basic (SW2), and PoE (SW3) switch.

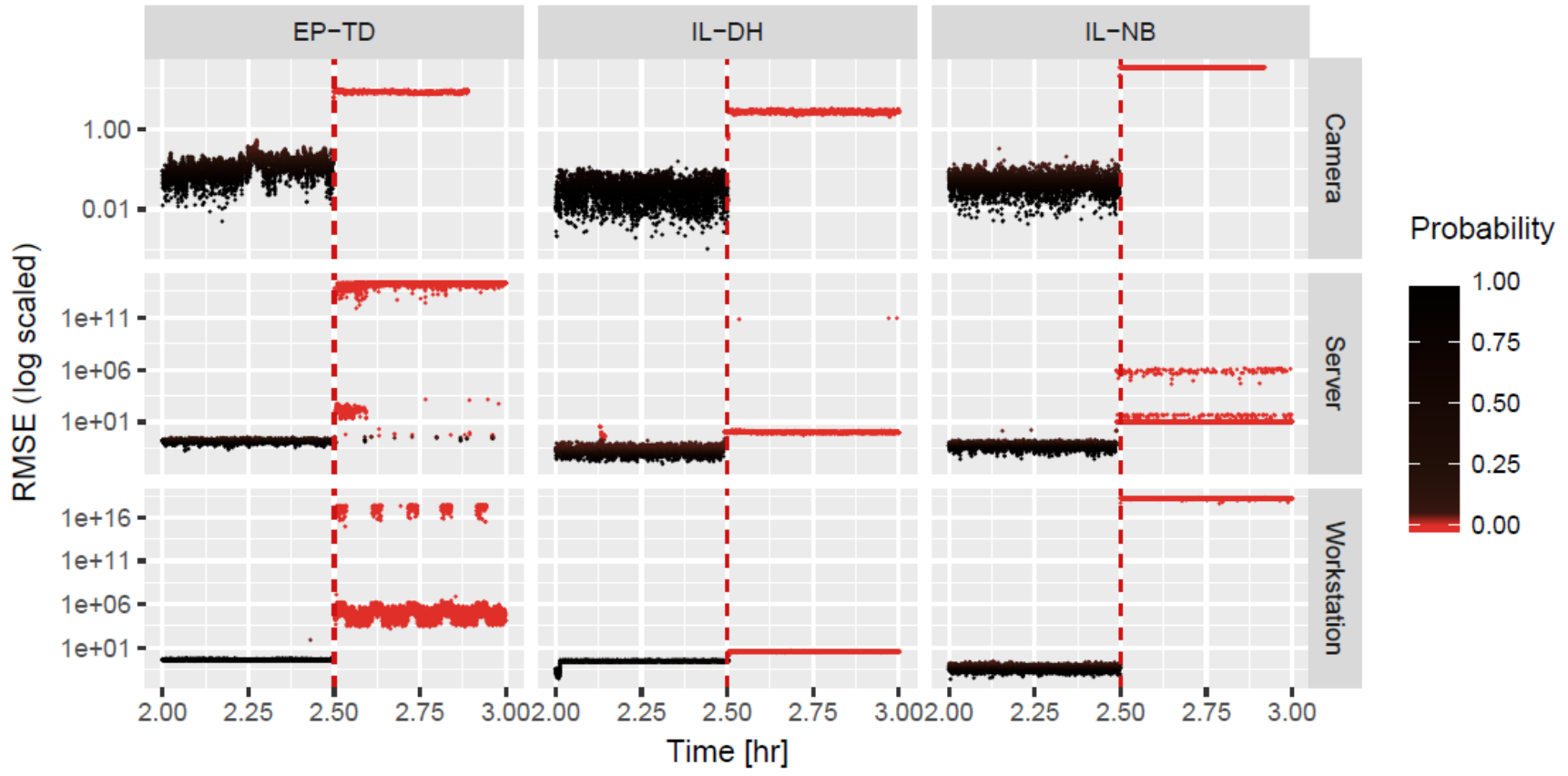

Fig. 12: Plots of the anomaly scores produced by *Vesper* when installed on a laptop PC (*Alice*), and in the presence of one intermediary switch. The rows indicate the victim (*Bob*), and the columns indicate the MitM attack (*Eve*).

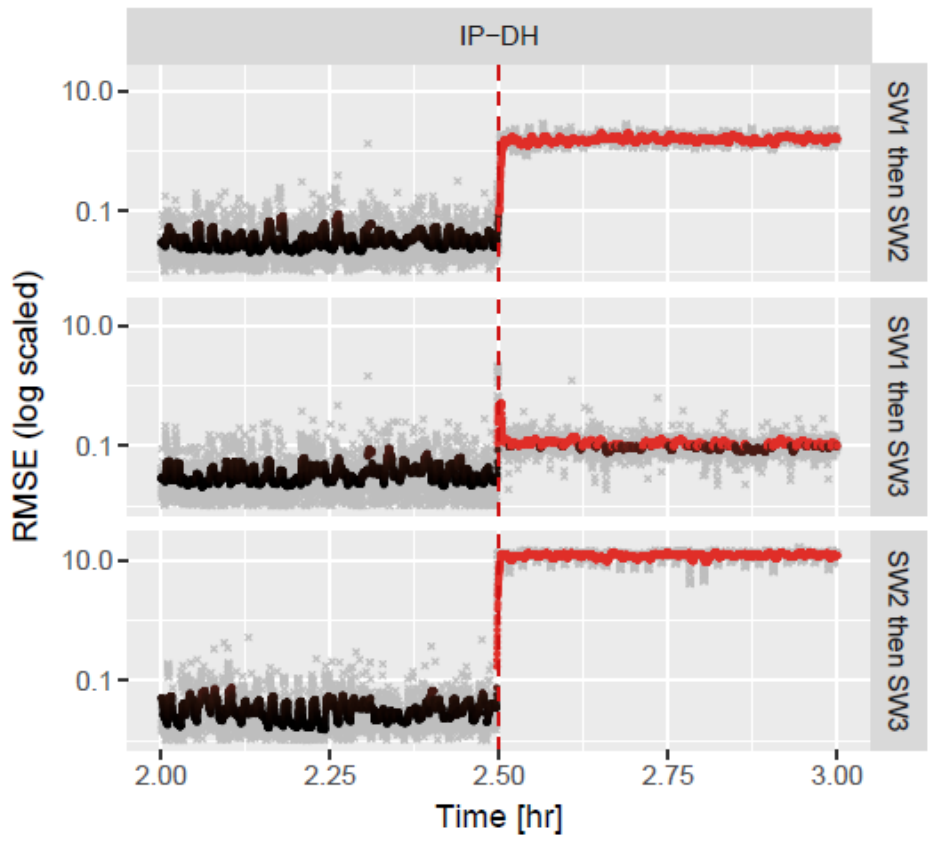

Fig. 13: *Vesper* detecting IP-DH MitM attacks. The scores before applying a 1 min averaging window are marked in gray.

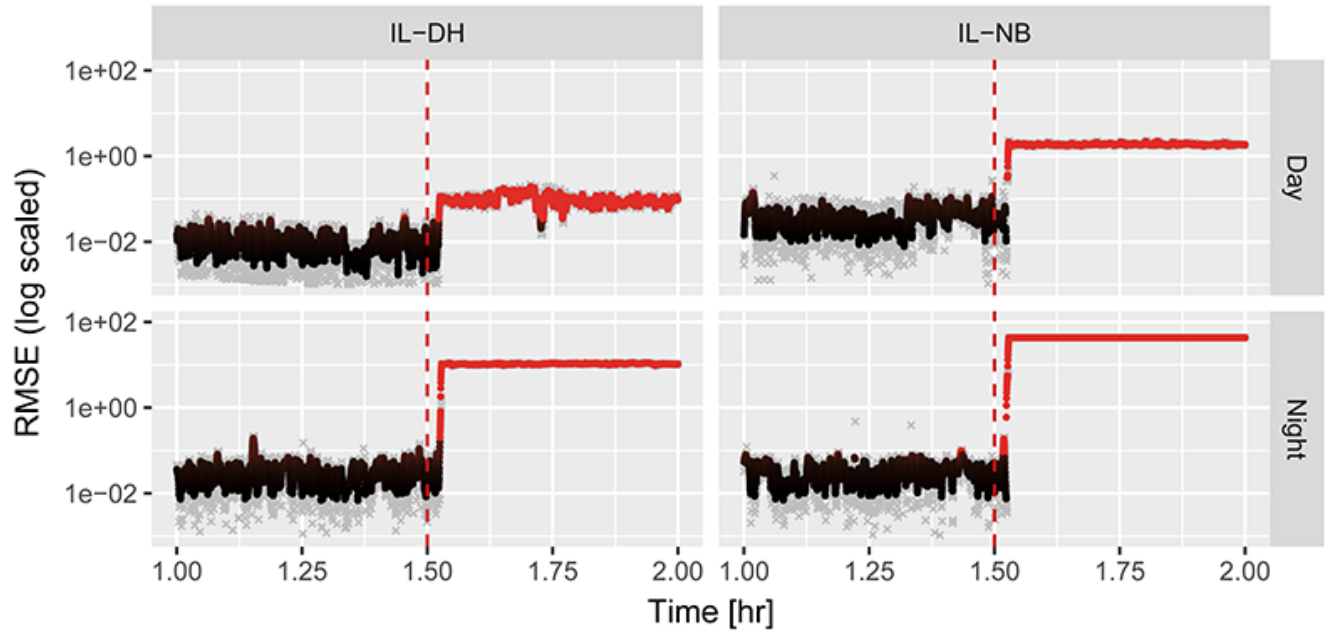

Fig. 14: *Vesper's* performance in detecting IL MitMs in a large LAN over multiple intermediary switches. The scores before applying 10 sec averaging window are marked in gray.

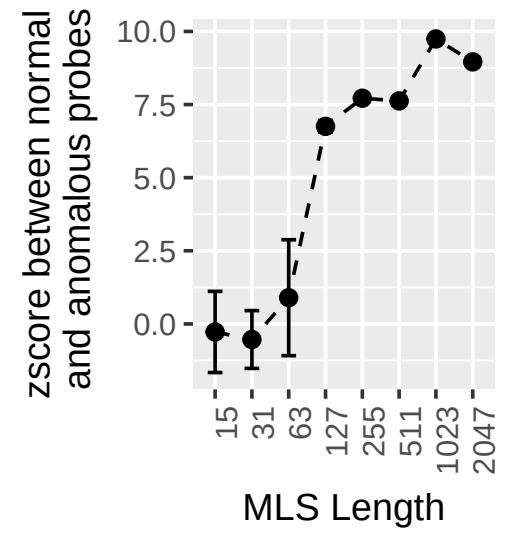

Fig. 15: How MLS signal size affects the detection of an IL MitM.

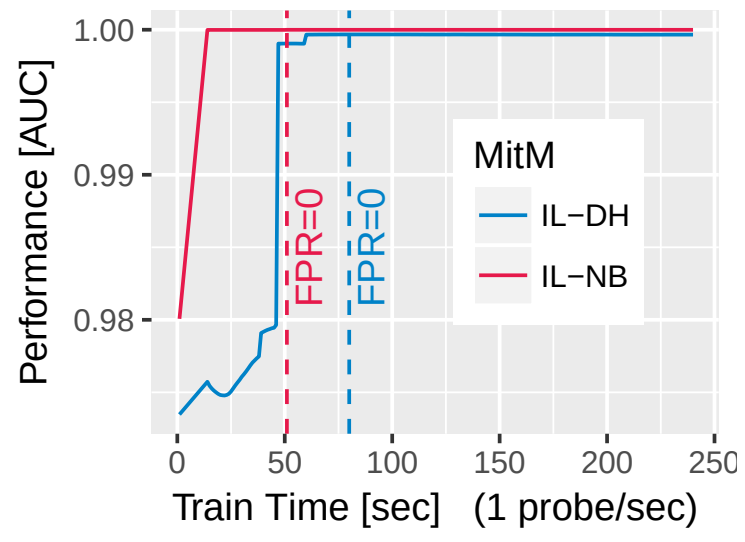

Fig. 16: The performance while retraining a model over multiple switches during the day.

In this section, we refer to each of the devices in the above setups as the attacker *Eve*. In our experiments, we used a C++ implementation of *Vesper*, set the autoencoder learning rate to $l = 0.1$, the *KS-Tester* parameters to $m = 5$ and $k = 15$, and the MLS probe length to $N = 1023$.

In order to evaluate *Vesper*'s accuracy when operating in different sized LANs, we examined two setups: (1) when *Alice* connects to the same switch as *Bob* (one intermediary switch), and (2) when *Alice* connects several switches away from *Bob* (multiple intermediary switches). In both setups, *Vesper* was evaluated in the presence of a wide variety of real-world traffic, while the end-hosts were actively using the network. We will now present our experimental results accordingly.

### A. One Intermediary Switch

For the EP and IL MitMs, we experimented on two LANs (1) a surveillance camera network, and (2) a LAN segment populated with active servers. The surveillance network consisted 8 high-end HD Sony cameras and three PCs. The server LAN segment consisted of one large switch connected to 61 active servers. *Alice* was a Kali Linux laptop PC (Intel i5 CPU), and *Bob* was either a camera (SNC-EB602R), a Windows desktop PC workstation (Intel i7 CPU), or a data server (Intel Xeon E5-2660 CPU) in each experiment. The desktop PC was located in the surveillance network.

We performed the EP-DH, IL-NB and IL-DH attacks on each of the three versions of *Bob*, with duration of 3 hours each. Fig. 12 shows RMSE scores produced by *Vesper* in each of experiments (at the moment of the attack). Each point in the figure represents the result of a single probe, and the color indicates the probe's abnormality.

For the IP-DH attack, we trained *Vesper* on one switch, and then swapped the switch with a different one. Fig. 13 presents the results from the IP-DH attacks.

In all experiments, the IL-NB was the easiest MitM attack to detect. This is because the Pi must perform additional logic in the kernel in order to bridge each ICMP packet. In contrast, the IL-DH and IP-DH were the most difficult to detect, because the packet interception was performed by dedicated hardware. We also note that there were 15 false positives in the experiment with the Server and the IL-DH due to a momentary disconnect. However, these FPs can be easily be mitigated by using an averaging window over the scores. This is because the mean of the scores' distribution significantly changes as seen in Fig.

TABLE II: A summary of this paper's experiments and results with regards to *Vesper*'s MitM detection performance.

| | **Figure:** | **12** | | | | | | | | | **13** | | | **14** | | | | **18** |
|---|---|---|---|---|---|---|---|---|---|---|---|---|---|---|---|---|---|---|
| **MitM Attack** | *Topology* | EP | IL | IL | EP | IL | IL | EP | IL | IL | IP | IP | IP | IL | IL | IL | IL | IL |
| | *Implementation* | TD | DH | NB | TD | DH | NB | TD | DH | NB | DH | DH | DH | NB | DH | NB | DH | DH |
| | *Alice* | Dell Laptop | | | Dell Laptop | | | Dell Laptop | | | Dell PC | | | Dell PC (Lab Workstation) | | | | Laptop |
| | *Bob* | SNC-EB602R Camera | | | Dell PC | | | Intel Xeon E5 Server | | | Dell Laptop | | | Dell PC (Secretary) | | | | PC |
| | *Eve* | Dell PC | Rasp. PI 3B | 1Gbps Switch | Dell PC | Rasp. PI 3B | 1Gbps Switch | Dell PC | Rasp. PI 3B | 1Gbps Switch | Basic Switch | PoE Switch | PoE Switch | Rasp. PI 3B | 1Gbps Switch | Rasp. PI 3B | 1Gbps Switch | 1Gbps Switch |
| | *Attack Vector* | Install Malw. | Add Device | Add Device | Install Malw. | Add Device | Add Device | Install Malw. | Add Device | Add Device | Replace Adva. | Replace Adva. | Replace Basic | Add Device | Add Device | Add Device | Add Device | Add Device |
| **Environment** | *Description* | Surveillance Network | | | Surveillance Network | | | Server Network | | | Basic LAN | | | Large Office Network | | | | Surv. |
| | *Num. Switches in the LAN* | 7 | 7 | 7 | 7 | 7 | 7 | 3 | 3 | 3 | 1 | 1 | 1 | 14 | 14 | 14 | 14 | 7 |
| | *Num. Intermediary Switches* | 1 | 1 | 1 | 1 | 1 | 1 | 1 | 1 | 1 | 1 | 1 | 1 | 4+ | 4+ | 4+ | 4+ | 4+ |
| | *Num. of hosts in the LAN* | 13 | 13 | 13 | 13 | 13 | 13 | 61 | 61 | 61 | 2 | 2 | 2 | 379 | 379 | 379 | 379 | 13 |
| | *Time of Day* | Day | Day | Day | Day | Day | Day | Day | Day | Day | Day | Day | Day | Day | Day | Night | Night | Day |
| **Experiment** | *Total Probes Sent* | 15,000 | 15,000 | 15,000 | 15,000 | 15,000 | 15,000 | 15,000 | 15,000 | 15,000 | 10,000 | 10,000 | 10,000 | 15,000 | 15,000 | 15,000 | 15,000 | 15,000 |
| | *Approximate Duration [hr]* | 4 | 4 | 4 | 4 | 4 | 4 | 4 | 4 | 4 | 3 | 3 | 3 | 4 | 4 | 4 | 4 | 4 |
| ***Vesper*'s Detection Performance** | *Area under the Curve (AUC)* | 1 | 0.9998 | 1 | 1 | 0.9997 | 1 | 0.9957 | 0.9952 | 0.9999 | 0.9999 | 0.9973 | 0.9999 | 0.9997 | 0.9978 | 0.9908 | 0.9992 | 0.9989 |
| | *Equal Error Rate (EER)* | 0 | 0.0001 | 0 | 0 | 0.0063 | 0 | 0.0088 | 0.0096 | 0.0008 | 0.0002 | 0.0170 | 0.0002 | 0.0004 | 0.0070 | 0.0089 | 0.0016 | 0.0203 |
| | *Recall (TPR)* | 1 | 0.9997 | 1 | 1 | 0.9931 | 1 | 0.9856 | 0.9984 | 0.9986 | 1 | 0.6536 | 1 | 1 | 0.9890 | 0.9935 | 1 | 0.8667 |
| | *False Alarm Rate* (FPR)* | 0 | 0.0001 | 0 | 0 | 0 | 0 | 0 | 0.0096 | 0 | 0.0002 | 0.0028 | 0.0002 | 0.0004 | 0.0040 | 0.0089 | 0.0016 | 0 |
| | *Accuracy* | 1 | 0.9998 | 1 | 1 | 0.9965 | 1 | 0.9928 | 0.9945 | 0.9993 | 0.9999 | 0.8250 | 0.9999 | 0.9998 | 0.9926 | 0.9922 | 0.9992 | 0.9327 |
| | *Precision* | 1 | 0.9999 | 1 | 1 | 1 | 1 | 1 | 0.9908 | 1 | 0.9998 | 0.9957 | 0.9998 | 0.9996 | 0.9956 | 0.9904 | 0.9982 | 1 |
| | *Detection Delay [sec]* | 0 | 0 | 0 | 0 | 0 | 0 | 0 | 0 | 0 | 0 | 0 | 0 | 0 | 0 | 0 | 0 | 8 |

**By using an averaging window of 60 seconds, all false positives in this table are mitigated. However, doing do incurs and additional delay in the detection time.*

12. However, the trade-off with using averaging window is that it causes a detection delay. Here, we found that a window size of one minute reduces the number of false positives to zero.

### B. Multiple Intermediary Switches

To evaluate *Vesper* across multiple intermediary switches, we used an organization's office LAN. The LAN consisted of over 379 network hosts connected through 14 large Ethernet switches, some of which used optical fiber uplinks. The hosts consisted of workstations, servers, printers, and IoT devices. The test scenario involved *Alice* (a desktop PC), and *Bob* (the secretary's PC) which were located on opposite sides of the LAN (separated by four large switches). The probes were sent for three hours, and the attacks (*Eve*) were IL MitMs only.

Fig. 14 shows *Vesper*'s RMSE scores during the day (busy hours) and during the night. The additional traffic during the day caused several false positives when detecting the IL-DH. However, by using an averaging window of 10 seconds, we were able to mitigate the errors completely. The results show that *Vesper* can detect IL MitMs sufficiently well in large noisey LANs, especially during off-hours.

### C. Profile Train Time

A concern with *Vesper* is that should a change occur in the LAN's topology, the affected models in the HP component must be retrained. When retrained during busy hours, in the event of an IL MitM, we found that *Vesper* reaches a false positive rate of zero within seconds (5-15 probes) when applied over one switch, and approximately a minute when applied across multiple switches (the large office LAN). Therefore, although *Vesper* is vulnerable during training, the attacker is challenged with deploying the MitM attack within a narrow time window. To make this window even smaller, *Vesper* can send probes at a faster rate during the grace period. Fig. 16 shows *Vesper*'s performance over time, in the case of multiple switches during day-time traffic. The performance was measured in AUC [28], interpretation: (1.0) *Vesper* was a perfect detector, (0.5) *Vesper* was randomly guessing.

### D. Probe Length

Fig. 15 shows how the parameter $N$ increases the separation between the normal probes and anomalous probes. Although the use of longer probes improves accuracy, there is a trade-off with bandwidth. *Vesper* sends one probe per second, and a probe has an average of $N(\frac{42+1542}{2})$ bytes. For example, with an $N = 1023$, the bandwidth used is approximately 810 Kbps. This rate is practical, especially since the probe traffic is contained within the LAN. However, a user should consider the number of *Vesper* instances installed to appropriately configure $N$ according to his/her limitations.

## VII. Adversarial Attacks

Our base assumption in this paper is that the MitM attack is introduced to the network after *Vesper*. However, even if Alice installs *Vesper* before Eve arrives, Eve can still attempt to evade detection.

We identify four possible adversarial attacks against *Vesper* (illustrated in Fig. 17): DoS, Spoof, Replay, and Bypass. *Vesper* can detect these evasions through the three features which the FE extracts from each probe. Each of the features is strong at detecting one particular attack, but weak at detecting another. However, when combined, the three features provide good protective coverage. Table III maps this relationship and provides a summary of each of the feature's strengths and weaknesses. We will now discuss the detection capabilities of each feature with respect to each evasion.

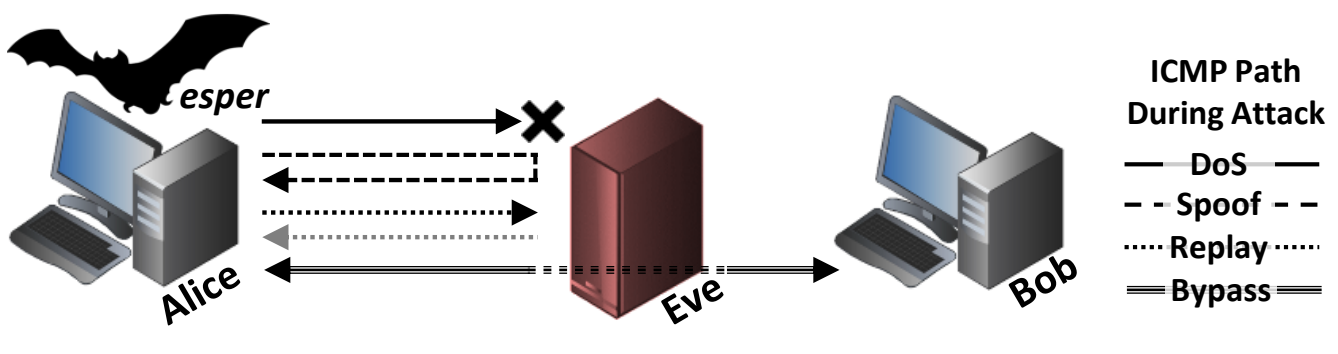

Fig. 17: The path of the ICMP packets sent from *Vesper* during each of the adversarial attacks.

TABLE III: The capabilities of each feature under each attack, and a summary of the features' strengths and weaknesses.

| Feature | DoS EP | DoS IL | DoS IP | Spoof EP | Spoof IL | Spoof IP | Replay EP | Replay IL | Replay IP | Bypass EP | Bypass IL | Bypass IP |
|---|---|---|---|---|---|---|---|---|---|---|---|---|
| $v_{E_h}$ | ● | ● | ● | ○ | ● | ● | ● | ● | ● | - | ○ | ○ |
| $v_{rtt*}$ | ● | ● | ● | ◌ | ○ | ◌ | ○ | ○ | ◌ | - | ○ | ◌ |
| $v_{jit}$ | ● | ● | ● | ● | ● | ● | ◌ | ◌ | ◌ | - | ○ | ○ |

**EP:** End-Point MitM
**IL:** In-Line MitM
**IP:** In-Point MitM

**Detection**
◌ *Weak*
○ *Modest*
● *Strong*

| Feature | Strengths | Weaknesses |
|---|---|---|
| $v_{E_h}$ | Detecting Replay Attacks | Has 1D Collision Space |
| $v_{rtt*}$ | Detecting Additional Hops | Detecting Spoof Attacks |
| $v_{jit}$ | Detecting Spoofing Attacks | Detecting Replay Attacks |

## A. *DoS*

In a DoS attack against *Vesper*, Eve drops `Echo_Request` packets *en route* to Bob so that Alice never receives the signal $y$. All features are strong against this attack because in the event of a loss-of-signal, a time-out occurs which causes a large spike in the features' values.

## B. *Spoof*

In a spoof attack against *Vesper*, Eve replies to `Echo_Request` packets *en route* to Bob, on behalf of Bob. In this way, no additional hops are added to the packet's path.

$v_{rtt*}$ is weak during a spoof attack by a EP MitM. This is because the average RTT remains statistically the same. However, $v_{rtt*}$ is modest against spoof attacks with an IL MitM because the affect on the feature depends on the location of the MitM (i.e., placing the IL MitM in front of Bob reduces the RTT time). Moreover, if both Alice and Bob have an instance of *Vesper*, then Eve will be detected by one of them.

Although the impulse response changes in the presence of a MitM, there is a possibility of collision in the feature space because we take the average of the signal's energy. Therefore, $v_{E_h}$ is modest against spoof attacks by an EP MitM. However, when launched from an IL MitM, $v_{E_h}$ is strong because the topological placement of the MitM highly affects $v_{E_h}$. $v_{jit}$ is strong against a spoof attack because the distribution of the jitter signal acts as a good fingerprint of the end-host's processing behaviors.

We note that the spoof attack is a difficult evasion to detect if (1) an EP MitM is used, and (2) the network topology Alice-Bob and Alice-Eve are identical. Regardless, even if Eve succeeds at replicating the exact hardware, firmware, and software of Bob's device, (1) Eve will only be able to attack the link with Bob, but not Carol (who has a different device), and (2) there are minute difference in Bob's hardware which truly give Bob's device a unique signature, thus making it difficult for Alice to spoof a reply from Bob.

Identical switches and end-devices obtain their 'fingerprints' from imperfections in the manufacturing process [29], [30]. *Vesper* captures these fingerprints collectively in $y$. To demonstrate the attacker's challenge in reproducing Bob's fingerprint, we performed the following experiment: (1) A set of network hosts $\Gamma$, from the same LAN, are selected. (2) *Vesper* (i.e., Alice) is trained to protect the link with host $i \in \Gamma$ (Bob). (3) After 2000 probes, host $j \in \Gamma$ (Eve) begins replying to *Vesper* instead of host $i$. (4) Steps 2-3 are repeated for every combination of $i, j \in \Gamma$. The pair $(i, j)$ is a spoof attack trial.

The above experiment was performed on two different LANs: $(\Gamma_{as})$ an assortment of 100 networked computers in a office LAN, and $(\Gamma_{pi})$ 46 Raspberry Pi 3B devices, all connected to a single Ethernet switch. The results are summarized in figures 20 and 21 in terms of AUC. The value of the AUC has the following interpretation: (1.0) *Vesper* was a perfect detector, (0.5) *Vesper* couldn't differentiate between hosts $i$ and $j$, and (0.0) *Vesper* thought that that host $j$ was host $i$.

The results show that *Vesper* is robust against spoof evasion attacks. Even if an attacker uses the same hardware/software as the victim, *Vesper* is likely to detect the change via the echo-analysis which it performs.

## C. *Replay*

In a replay attack against *Vesper*, whenever a probe $x$ is intercepted on its way to Bob, Eve replays a previously intercepted response signal $y$ back to Alice.

$v_{E_h}$ is strong against all replay attacks because it is dependent on the MLS signal, and the MLS sequence is difficult to predict in real-time.[2] We note that the duration of a signal $x$ is not a constant and can be very noisy due to Alice's host's scheduler. This strengthens the detection capabilities of $v_{E_h}$ and $v_{rtt*}$ in the case of a replay attack. This is because the noise adds a nondeterministic skew to the $tx$ times, which Eve cannot predetermine. However, since Alice can unintentionally mitigate the noise by using a dedicated hardware/software, we consider $v_{rtt*}$ to be modestly secure against such attacks.

The $v_{jit}$ feature is very weak to replay attacks since the feature is not dependent on the uniqueness of the MLS signal. In Fig. 19, we present the affect a replay attack has on each of the features (top), and show that the evasion is successfully detected via the final computed anomaly score (bottom).

## D. *Bypass*

In a bypass attack against *Vesper*, an advanced attacker uses a special IL device which can choose to either (1) interact with the network acting as a MitM (active-mode), or (2) passively observe acting as a wiretap (passive-mode). To evade detection, Eve is either (A) always in active-mode and switches to passive-mode when an ICMP request is received, OR (B) always in passive-mode and switches to active-mode only whenever Eve wants to manipulate or inject traffic. *Vesper* can only detect Eve while she is in active-mode.

*Vesper* can detect Eve if she uses (A). This is because, by the time the first ICMP packet in $x$ is detected by Eve, the frame has already been partially buffered. Therefore, Eve must pass $x[1]$ through her regular interception process before switching over to passive-mode (see Fig. 20 for results). Furthermore, if Eve uses (B), then it is likely that *Vesper* will detect her. This is because Eve must remain in active-mode for long durations in order to be effective. For example, to manipulate

[2] This is true if each subsequent MLS seed is determined by a secure pseudo random number generator, such as AES-256 in CTR-mode.

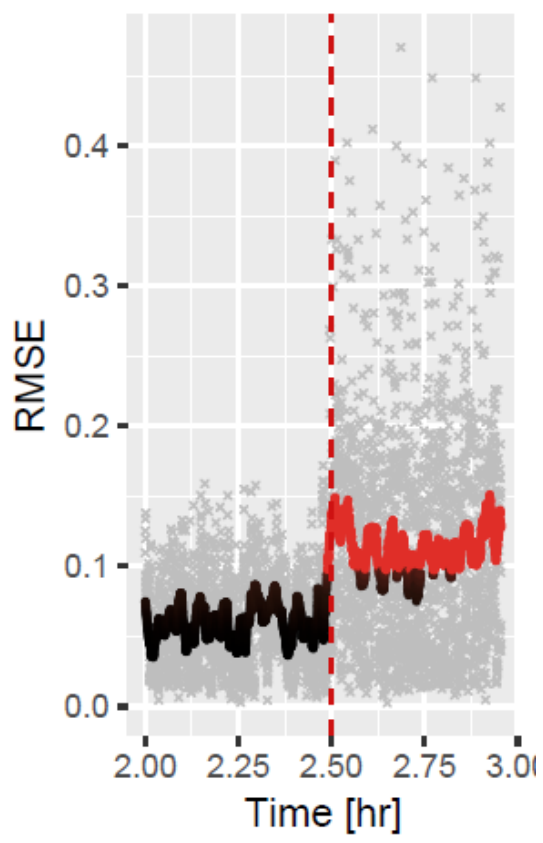

Fig. 18: *Vesper*'s detection of a IL-NB MitM using bypass evasion.

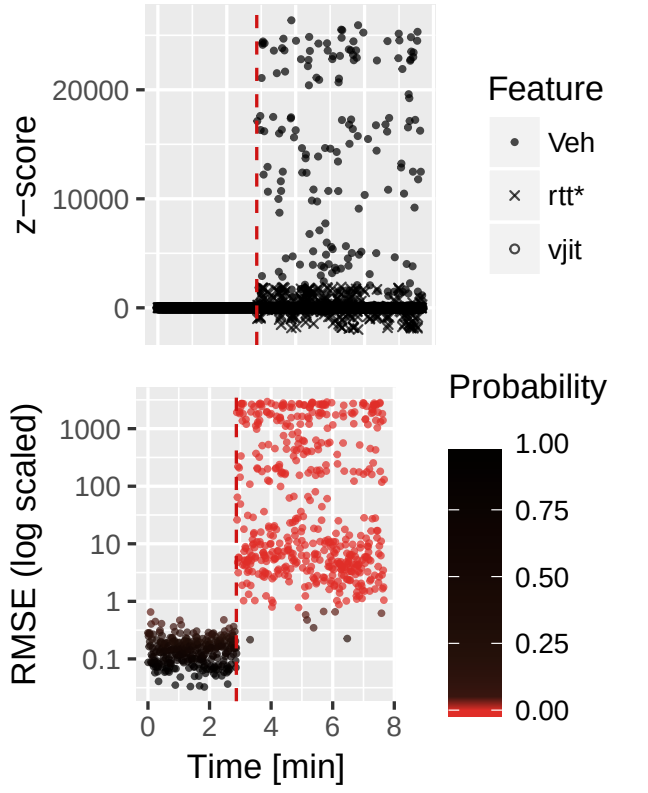

Fig. 19: *Vesper*'s detection of a replay attack. Top: the features in $\vec{v}$. Bottom: The autoencoder's anomaly scores.

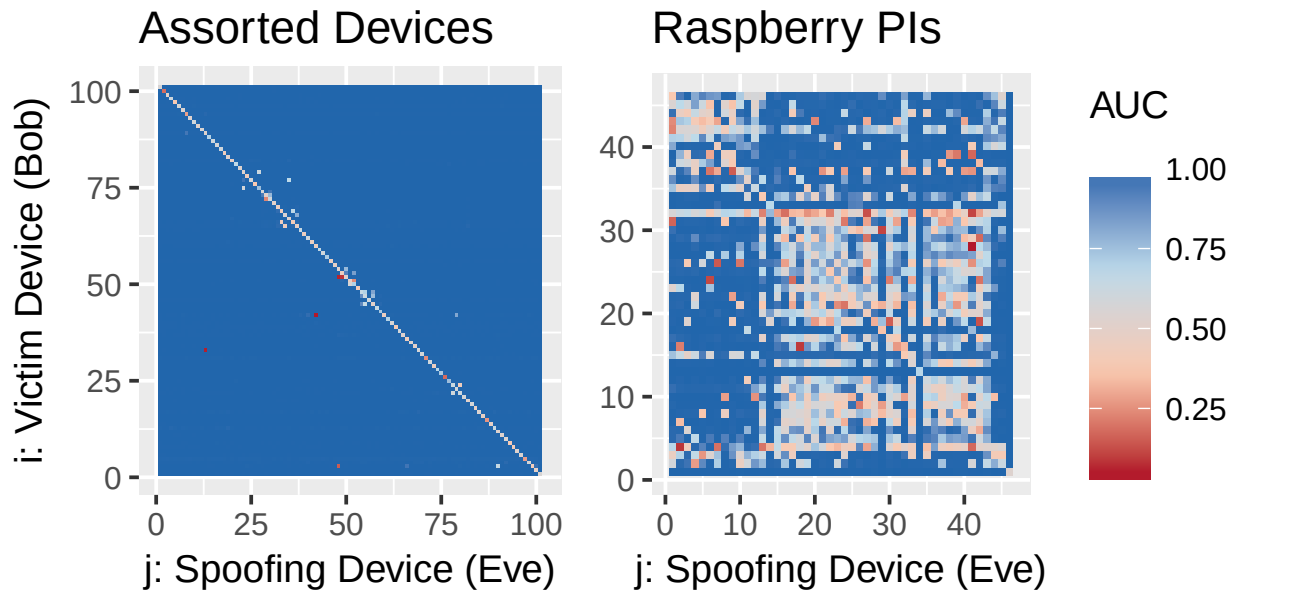

Fig. 20: *Vesper*'s AUC in detecting 12,116 different spoof attack trials in a CM setup, when Eve uses an arbitrary device (left), or a similar device (right).

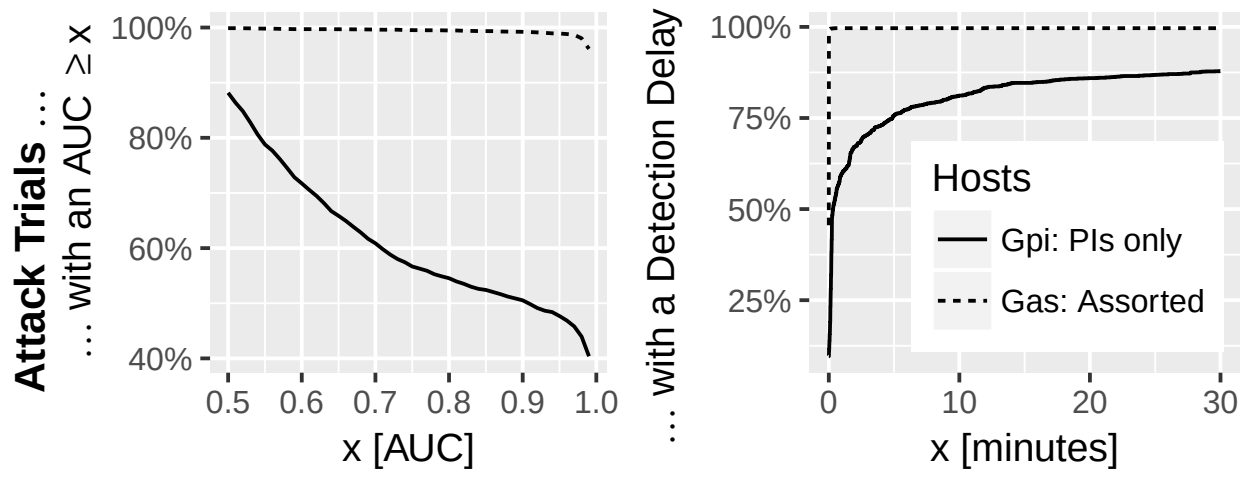

Fig. 21: The AUC and detection delays of the spoof attack experiments in terms of percentiles.

streaming/live data, maintain a compromised TCP connection, or to intercept a choice packet.

## VIII. Conclusion

As of today, MitM attacks are still pose a great threat to many LANs. In this paper we have presented a new technique for detecting MitM attacks in LANs via ping echo analysis. We have shown how the technique can be practically applied via a MitM detection framework called *Vesper*. Experimental results show that (1) *Vesper* is capable of detecting end-point, in-line, and in-point MitM attacks, and (2) is robust against possible adversarial attacks. For future work, we plan on applying other ping methods (e.g., TCP SYN), applying noise mitigation techniques, and extending the technique to work over routers, and applying *Vesper* to Wi-Fi networks.

## APPENDIX A

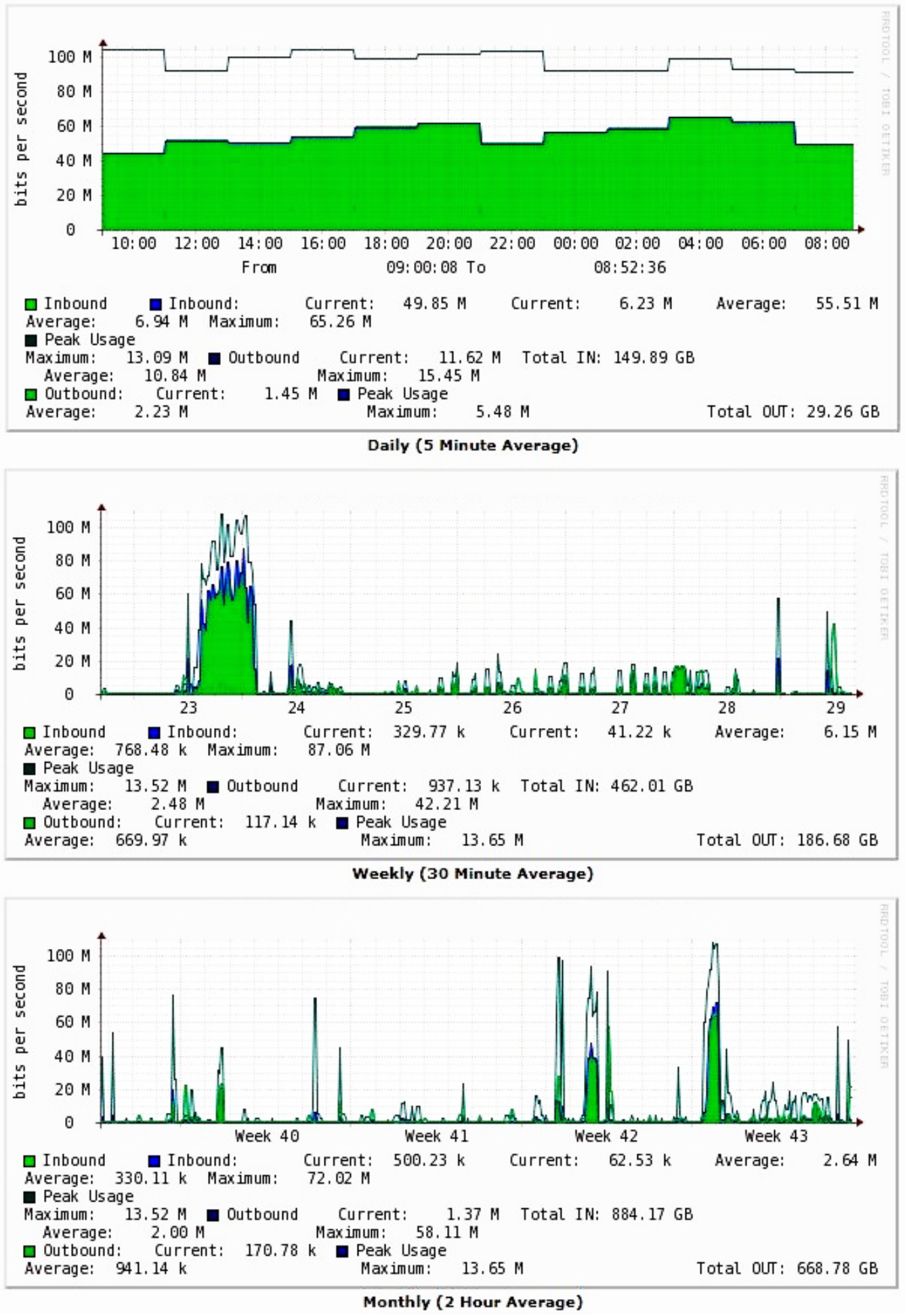

Fig. 22: The common hourly, daily, and monthly traffic loads of the 50 port switch used in the detection experiment where the Data Server was the victim (see Fig. 12).

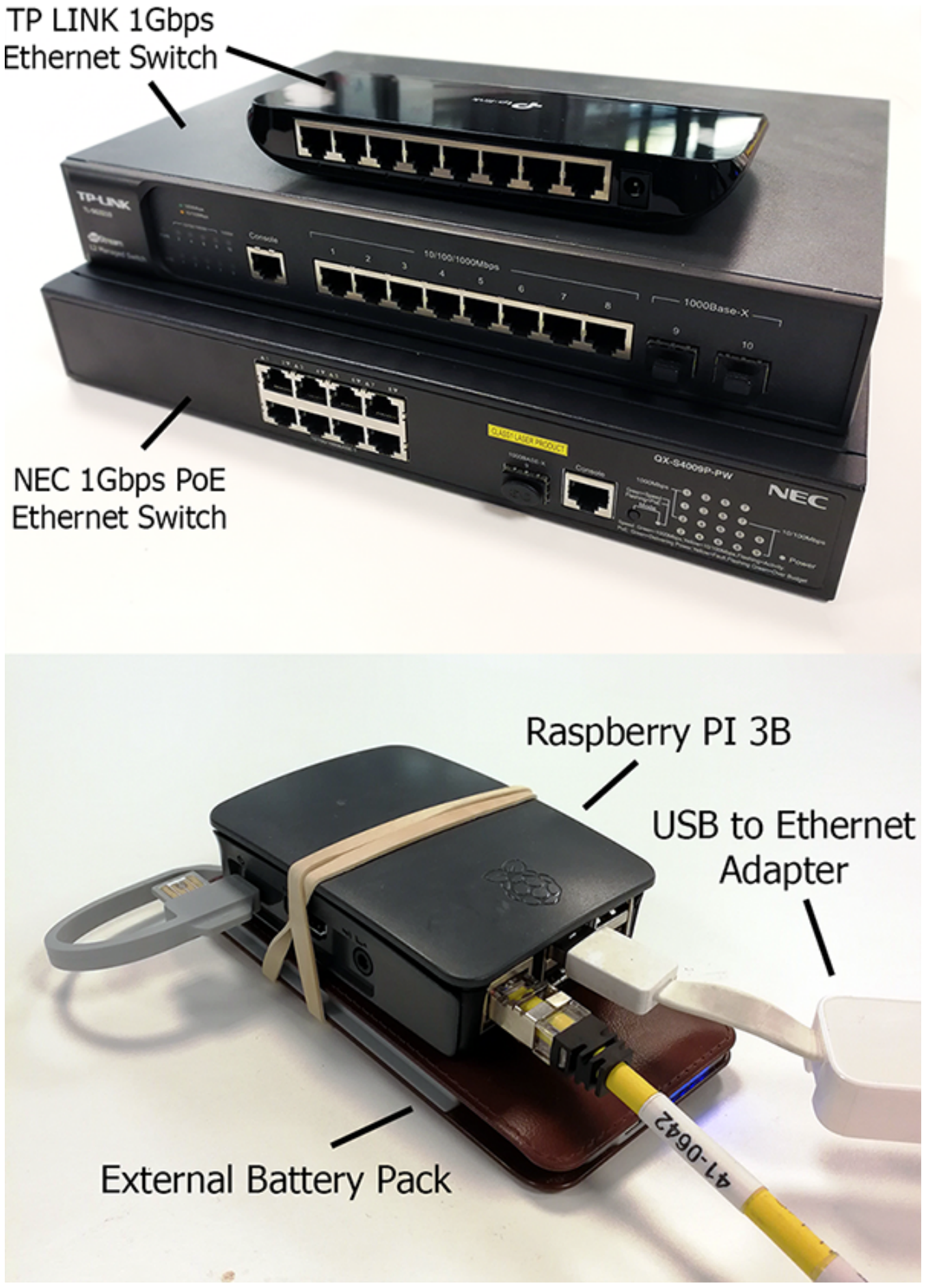

Fig. 23: The MitM devices used in this paper. Top: Three 1Gbps Ethernet switches used in the IP-DH experiments, where the middle switch was used in the IL-DH experiments. Bottom: A Raspberry Pi 3B with a battery pack and 1Gbps USB to Ethernet adapter, used in the IL-NB experiments.

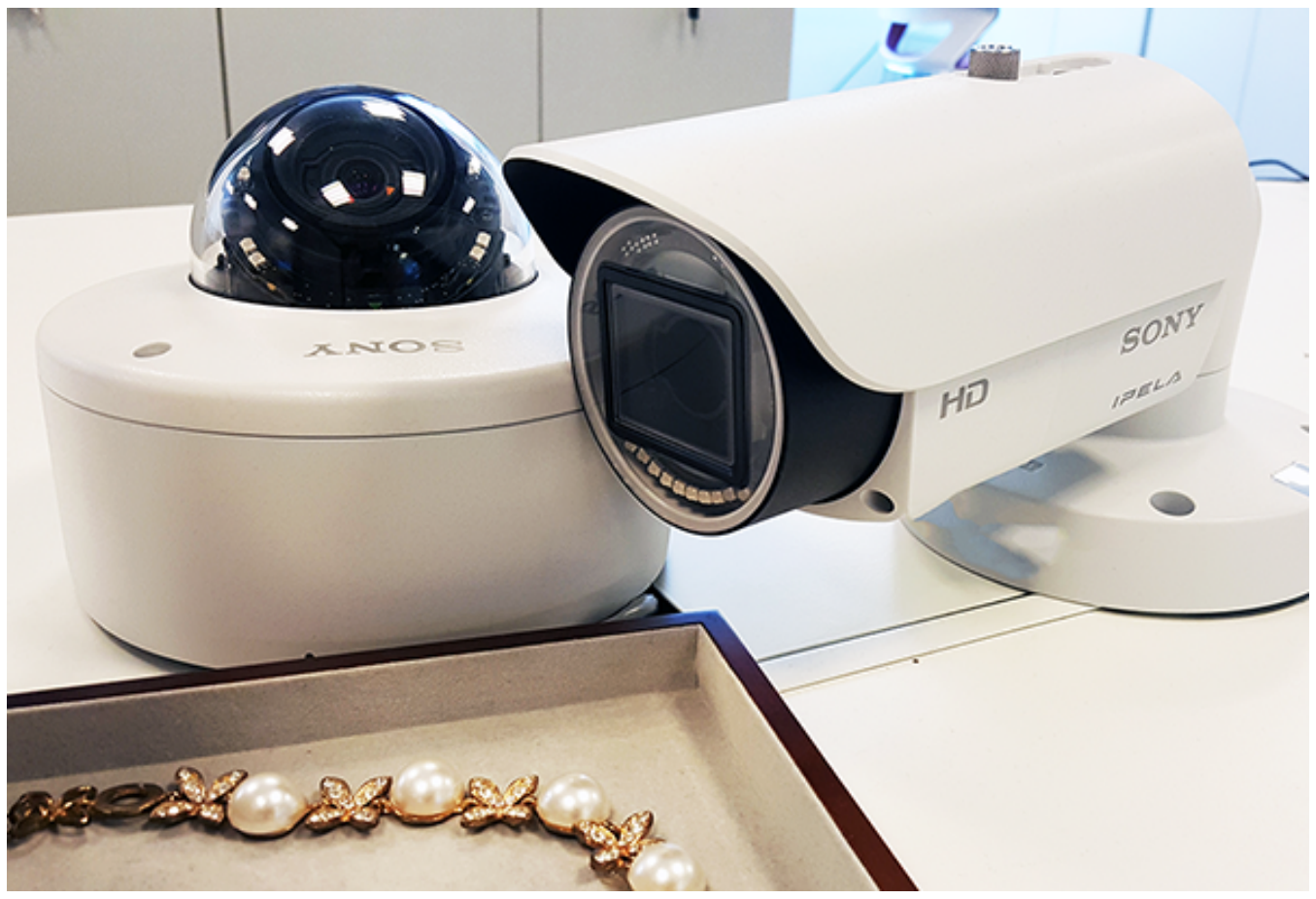

Fig. 24: Two of the eight Sony IP surveillance cameras used in the experiments. The models were: SNC-EM602RC, SNC-EB600, SNC-EM600, and SNC-EB602R.

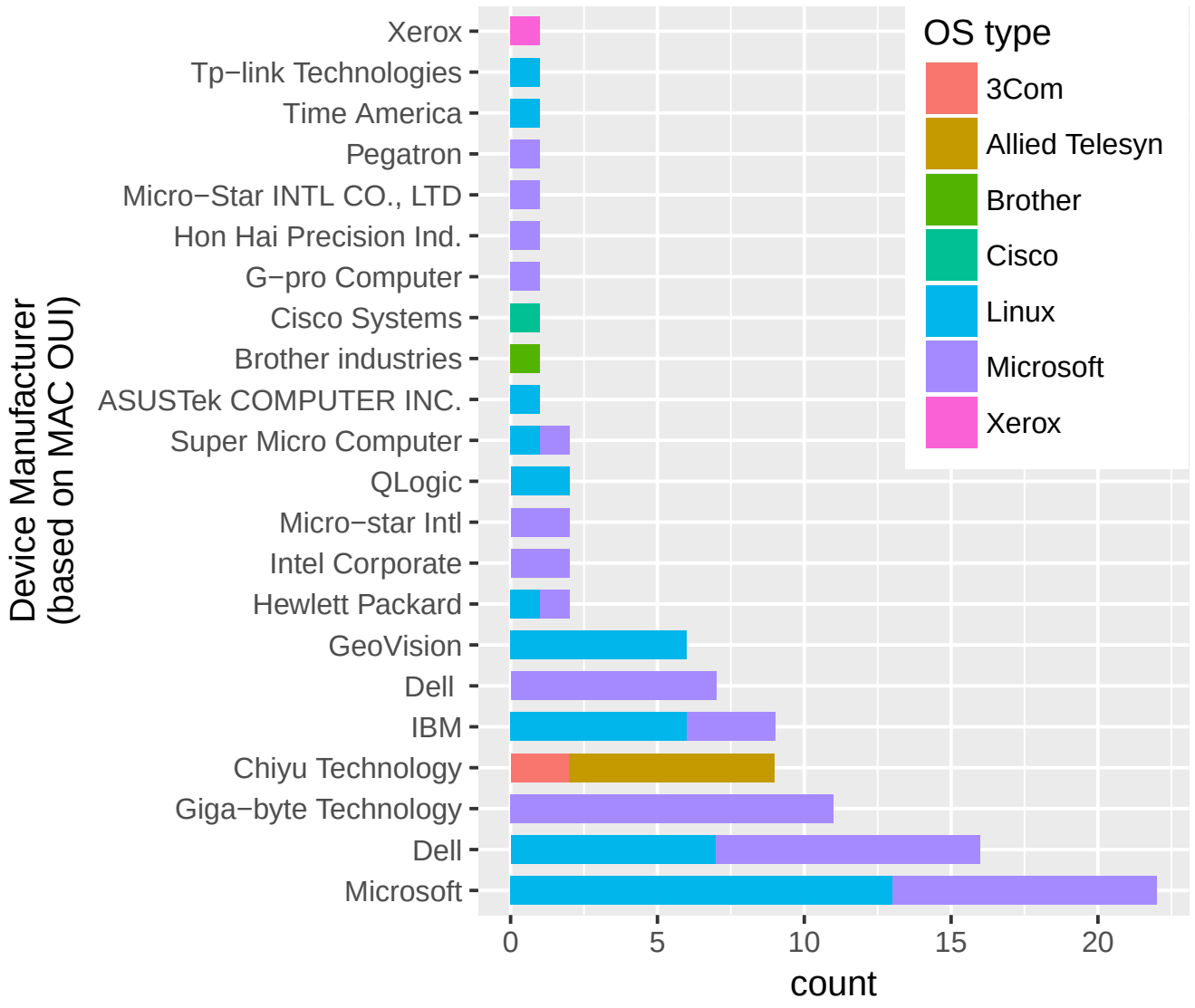

Fig. 25: A break-down of the device's operating systems, used in 100 host spoof experiment.

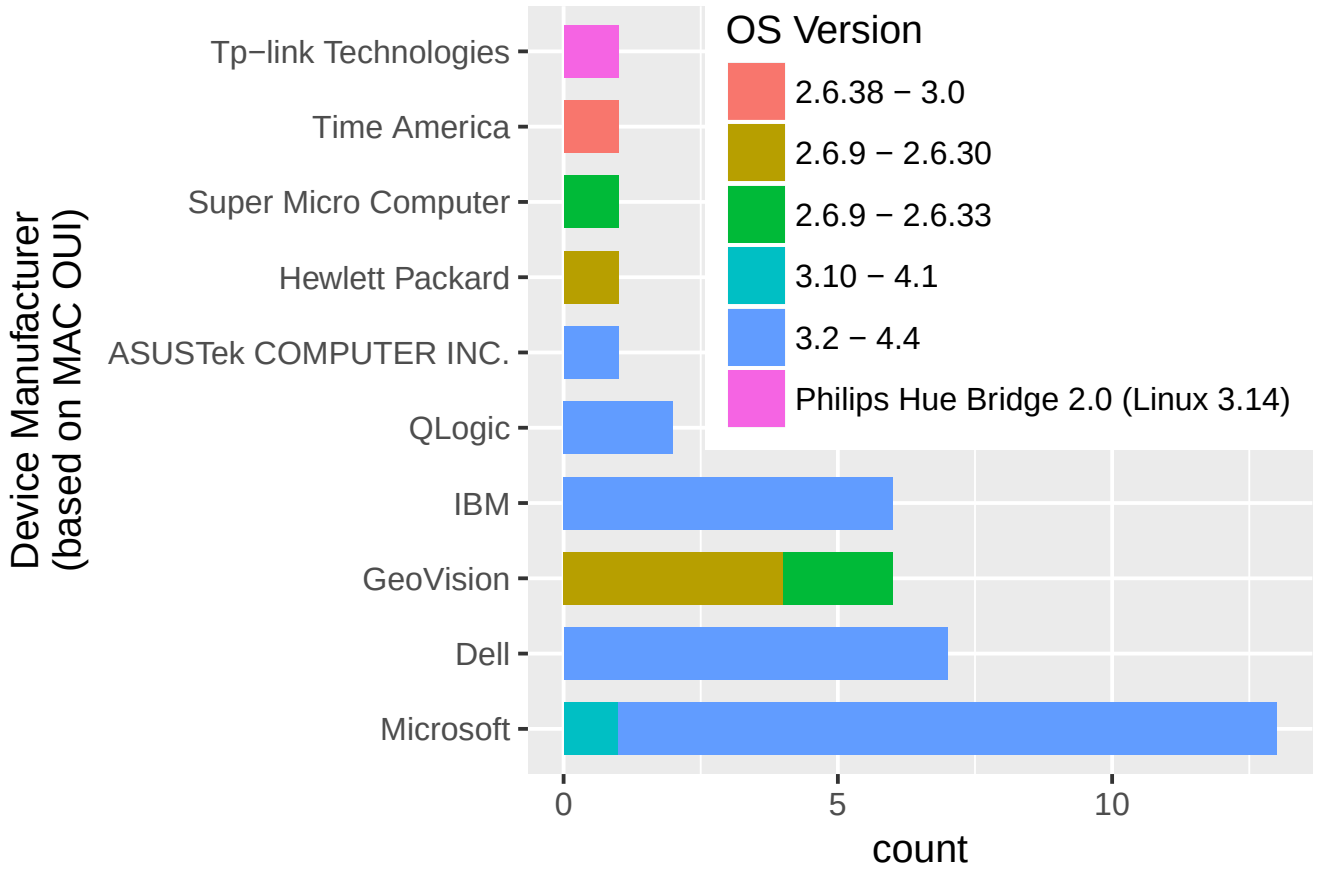

Fig. 26: A break-down of the devices running a Linux operating system, used in 100 host spoof experiment.

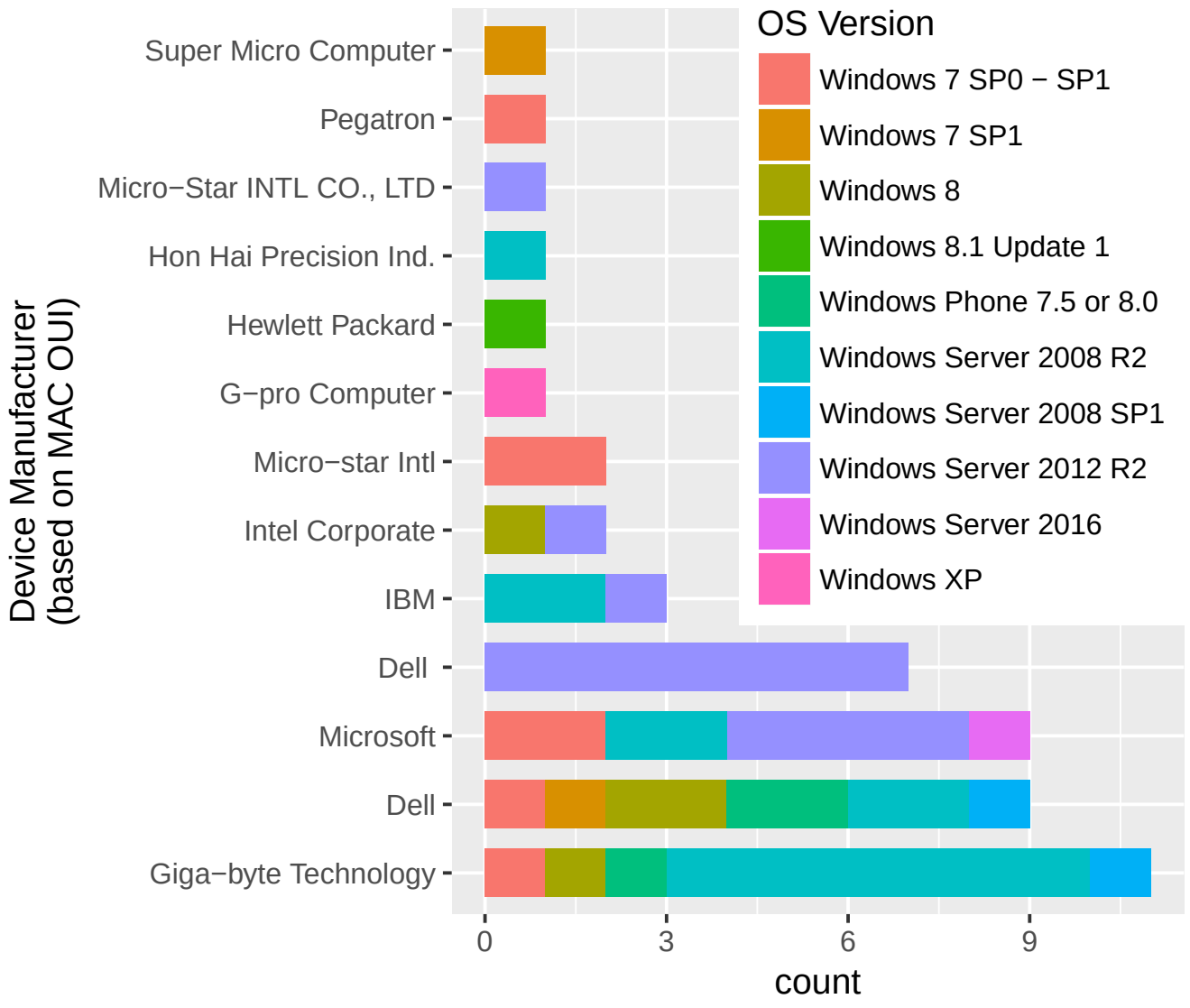

Fig. 27: A break-down of the devices running a Microsoft operating system, used in 100 host spoof experiment.

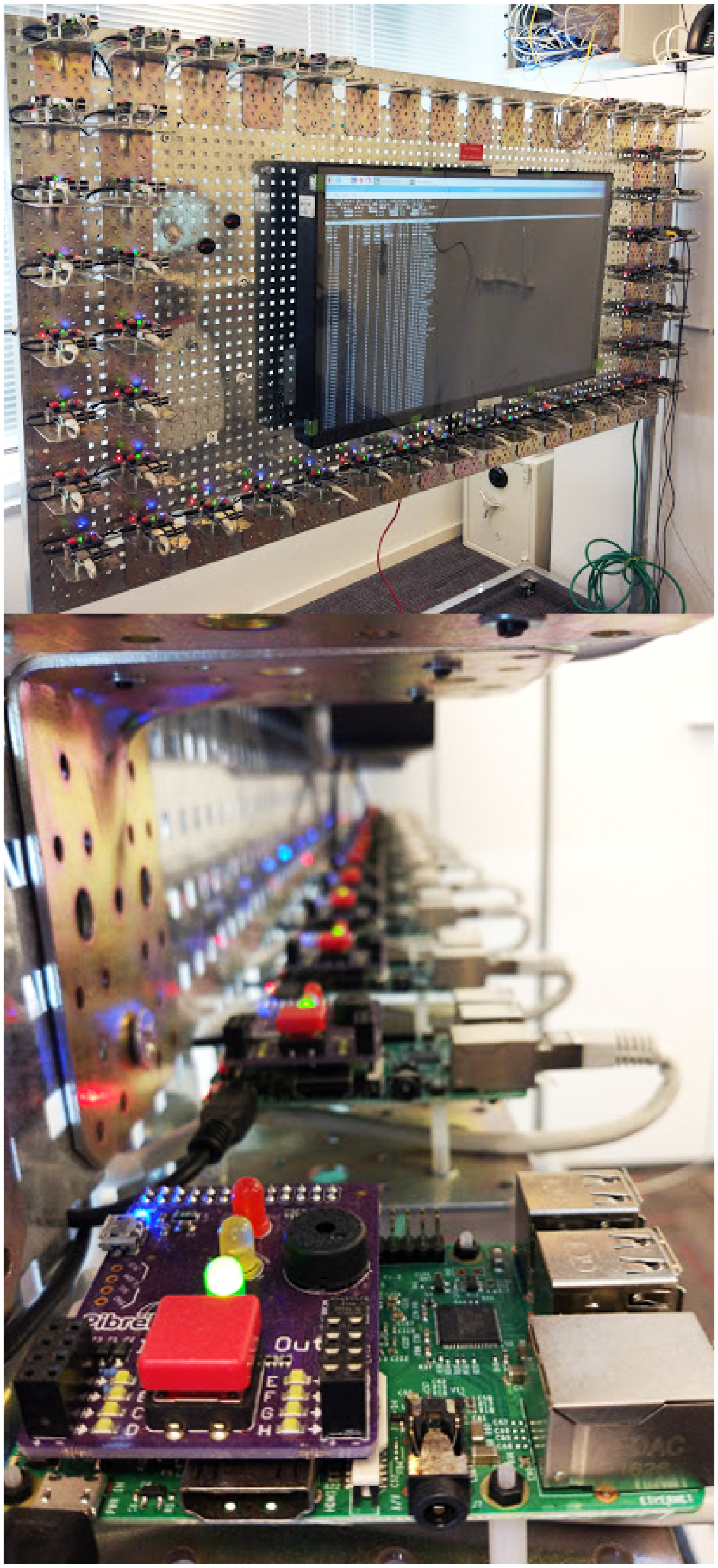

Fig. 28: The 46 Raspberry Pi 3Bs used in the experiments. All of the devices were connected to a single Ethernet switch.

```
1   #include "mls.h"
2
3   //local members:Taps,nbits
4   mls::mls(int n_bits)
5   {
6   if(n_bits<2)
7       nbits=2;
8   else if(n_bits>32)
9       nbits=32;
10  else
11      nbits=n_bits;
12  Taps[2]={1};Taps[3]={2};Taps[4]={3};Taps[5]={3};Taps[6]={5};Taps[7]={6};Taps[8]={7,6,1};Taps[9]={5};Taps[10]={7};Taps[11]={9};Taps[12]={11,10,4};Taps[13]={12,11,8};Taps[14]={13,12,2};Taps[15]={14};Taps[16]={15,13,4};Taps[17]={14};Taps[18]={11};Taps[19]={18,17,14};Taps[20]={17};Taps[21]={19};Taps[22]={21};Taps[23]={18};Taps[24]={23,22,17};Taps[25]={22};Taps[26]={25,24,20};Taps[27]={26,25,22};Taps[28]={25};Taps[29]={27};Taps[30]={29,28,7};Taps[31]={28};Taps[32]={31,30,10};
13  srand(time(NULL));
14  }
15
16  vector<bool> mls::get_seq()
17  {
18      vector<int> taps = Taps[nbits];
19      vector<bool> state(nbits);
20
21      uint32_t randomNum = rand();
22      for (int i = 0;i< nbits;i++)
23          state[i] = (randomNum >> i) & 1;
24
25      int length=pow(2,9) - 1;
26      vector<bool> seq(length);
27      bool feedback;
28
29      int idx = 0;
30
31      for( int i = 0; i< length; i++)
32      {
33          feedback = state[idx];
34          seq[i] = feedback;
35          for(int ti = 0; ti < taps.size();ti++)
36              feedback ^= state[(taps[ti] + idx) % nbits];
37          state[idx] = feedback;
38          idx = (idx + 1) % nbits;
39      }
40      return seq;
41  }
```

Fig. 29: The MLS generator used in the experiments. Note, on line 21, the random number should be drawn from a secure PRNG (e.g., AES-256 in CTR-mode).

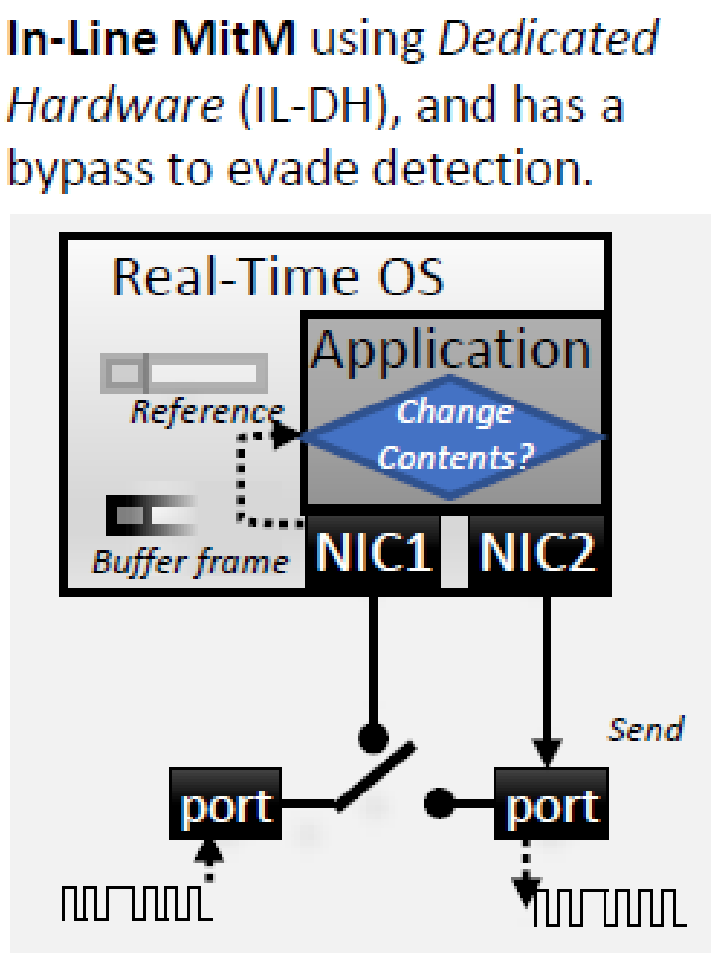

Fig. 30: The packet interception process of an IL-DH MitM device, with a bypass evasion mechanism. The device can either actively or passively observe traffic.

## APPENDIX B

In this section we show that there is a dependency between the the $k$-th RTT in $y$ to the payload $k$ payload sizes of the ICMP request packets sent prior in $x$. In our experiment, we used two Desktop PCs running Kali Linux (denoted PC1 and PC2), and a 1 Gbps Ethernet switch (SW). PC1 sent $20,000$ MLS input signals ($X = \{x_1, x_2, \ldots x_{20,000}\}$) and received the respective $20,000$ output signals ($Y = \{y_1, y_2, \ldots\}$). The transmission rate of the ICMP packets in each input signal was

$$f_s = \frac{2}{\mu_{RTT^*}} = \frac{2}{0.0001763151} = 11343.33Hz \qquad (13)$$

With the collected input and output signals, we built the linear regression model $y[25] \sim x[1], x[2], \ldots, x[25]$. In other words, the model predicts $y_i[25]$ given the descriptors $x_i[1], x_i[2], \ldots x_i[25]$ for some $x_i \in X$ and $y_i \in Y$. The p-value of each descriptor variable is available in Table IV.

TABLE IV: The linear dependencies of descriptor variables in the regression model $y[25] \sim x[1], x[2], \ldots, x[25]$.

| Descriptor | t value | $Pr(>\|t\|)$ | Signif. |
|---|---|---|---|
| *[Intercept]* | 255.59 | < 2e-16 | *** |
| $x[1]$ | -8.145 | 4.02E-16 | *** |
| $x[2]$ | 1.78 | 0.075028 | . |
| $x[3]$ | 1.632 | 0.102728 | |
| $x[4]$ | 1.685 | 0.092036 | . |
| $x[5]$ | -2.381 | 0.017295 | * |
| $x[6]$ | 1.056 | 0.290963 | |
| $x[7]$ | 1.576 | 0.114999 | |
| $x[8]$ | -1.092 | 0.274639 | |
| $x[9]$ | -1.875 | 0.060757 | . |
| $x[10]$ | -1.168 | 0.243001 | |
| $x[11]$ | -3.64 | 0.000273 | *** |
| $x[12]$ | 1.583 | 0.113487 | |
| $x[13]$ | -4.689 | 2.76E-06 | *** |
| $x[14]$ | 8.743 | < 2e-16 | *** |
| $x[15]$ | -19.653 | < 2e-16 | *** |
| $x[16]$ | 9.188 | < 2e-16 | *** |
| $x[17]$ | -1.976 | 0.048148 | * |
| $x[18]$ | 1.084 | 0.278364 | |
| $x[19]$ | -1.786 | 0.074057 | . |
| $x[20]$ | 3.5 | 0.000466 | *** |
| $x[21]$ | -6.475 | 9.72E-11 | *** |
| $x[22]$ | -7.038 | 2.01E-12 | *** |
| $x[23]$ | 23.562 | < 2e-16 | *** |
| $x[24]$ | -88.404 | < 2e-16 | *** |
| $x[25]$ | 258.505 | < 2e-16 | *** |

**Signif. codes:**
0 ***
0.001 **
0.01 *
0.05 .
0.1
1

The results demonstrate that the $k$-th RTT in the output signal $y$ is highly dependent on the $k$ ICMP payload sizes sent prior. We note that each input signal in $X$ was a completely random MLS sequence. Therefore, we conclude that the observed statistical dependency is the result of a dynamic reaction which SW and PC2 have to an MLS ping input sequence. This indicates that the output signal $y$ indeed captures a representation